\documentclass[fleqn,usenatbib]{mnras}

\usepackage{newtxtext,newtxmath}

\usepackage[T1]{fontenc}

\DeclareRobustCommand{\VAN}[3]{#2}
\let\VANthebibliography\thebibliography
\def\thebibliography{\DeclareRobustCommand{\VAN}[3]{##3}\VANthebibliography}

\usepackage{graphicx}	
\usepackage{amsmath}	
\usepackage{overpic}
\usepackage{pdflscape}  
\usepackage{adjustbox}

\usepackage[nolist]{acronym}
\usepackage[noabbrev]{cleveref}
\usepackage[range-units = single]{siunitx}
\usepackage{booktabs}
\usepackage{bm}

\usepackage{xcolor} 

\defcitealias{Dorsman2025}{D25}
\defcitealias{Salmi2018}{S18}

\newcommand\disc{\texttt{Disc}}
\newcommand\discline{\texttt{Disc+Line}}
\newcommand\st{\ac{ST}} 
\newcommand\stu{\ac{STU}}
\newcommand\std{\texttt{ST-Disc}}
\newcommand\stdl{\texttt{ST-DiscLine}}
\newcommand\stm{\texttt{ST-Marg}}
\newcommand\stud{\texttt{STU-Disc}}

\newcommand\stum{\texttt{STU-Marg}}
\newcommand\stume{\texttt{STU-MargEoS}}

\acrodefplural{EoS}{equations of state}
\acrodefplural{LMXB}{low-mass X-ray binaries}

\begin{acronym}
\acro{1D}{one dimensional}
\acro{2D}{two dimensional}
\acro{3D}{three dimensional}
\acro{4D}{four dimensional}
\acro{5D}{five dimensional}
\acro{AMP}{accreting millisecond pulsar}
\acro{CI}{credible interval}
\acro{EoS}{equation of state}
\acro{GR}{general relativistic}
\acro{MHD}{magneto-hydrodynamic}
\acro{GW}{gravitational wave}
\acro{HPC}{high-performance computing}
\acro{ISM}{interstellar medium}
\acro{IXPE}{Imaging X-ray Polarimetry Explorer}
\acro{J1808}{SAX J1808.4$-$3658}
\acro{LMXB}{low mass X-ray binary}
\acro{MP}{millisecond X-ray pulsar}
\acro{MAP}{maximum a posteriori}
\acro{NICER}{Neutron Star Interior Composition Explorer}
\acro{NS}{neutron star}
\acro{PPD}{posterior probability distribution}
\acro{PPM}{Pulse-profile Modelling}
\acro{RMP}{rotation powered millisecond pulsar}
\acro{RXTE}{{\it Rossi X-ray Timing Explorer}}
\acro{ST}[\texttt{ST}]{Single Temperature}
\acro{STU}[\texttt{ST-U}]{Single Temperature - Unshared}
\acro{TBO}{thermonuclear burst oscillation}
\acro{QPO}{quasi-periodic oscillation}
\acro{X-PSI}[\textsc{X-PSI}]{X-ray Pulse Simulation and Inference}
\end{acronym}

\crefformat{equation}{equation~#2#1#3}
\newcommand\T{\rule{0pt}{2.8ex}}       
\newcommand\B{\rule[-1.6ex]{0pt}{0pt}} 

\title[Pulse profile modelling of SAX J1808]{Pulse profile modelling of the accretion-powered millisecond pulsar SAX J1808.4$-$3658 using NICER data from its 2019 and 2022 outbursts}

\author[Dorsman et al.]{Bas Dorsman,$^{1}$\thanks{E-mail: b.dorsman@uva.nl} 
Tuomo~Salmi,$^{2}$
Anna~L.~Watts,$^{1}$
Mason Ng,$^{3,4,5}$
Anna Bobrikova,$^{6}$
Vladislav Loktev,$^{2,6}$\newauthor
Juri Poutanen,$^{6}$
and Joern Wilms$^{7}$
\\
$^{1}$Anton Pannekoek Institute for Astronomy, University of Amsterdam, Science Park 904, 1098XH Amsterdam, the Netherlands\\
$^{2}$Department of Physics, P.O. Box 64, FI-00014 University of Helsinki, Finland\\
$^{3}$MIT Kavli Institute for Astrophysics and Space Research, Massachusetts Institute of Technology, 77 Massachusetts Avenue, Cambridge, MA 02139, USA\\
$^{4}$Department of Physics, McGill University, 3600 rue University, Montr\'{e}al, QC H3A 2T8, Canada\\
$^{5}$Trottier Space Institute, McGill University, 3550 rue University, Montr\'{e}al, QC H3A 2A7, Canada\\
$^{6}$Department of Physics and Astronomy, FI-20014 University of Turku, Finland\\
$^{7}$Dr. Karl Remeis-Observatory and Erlangen Centre for Astroparticle Physics, Universität Erlangen-Nürnberg, Sternwartstr. 7,
96049 Bamberg, Germany \\
}

\date{Accepted XXX. Received YYY; in original form ZZZ}

\pubyear{\the\year{}}

\begin{document}
\label{firstpage}
\pagerange{\pageref{firstpage}--\pageref{lastpage}}
\maketitle

\begin{abstract}
Pulse profile modelling is a relativistic ray-tracing technique that has provided constraints on parameters, with a focus on mass and radius, of five rotation-powered millisecond pulsars. While the technique can also be applied to accretion-powered millisecond pulsars (AMPs), this requires accounting for the X-rays from the accretion disc and has only been applied to archival data from the Rossi X-ray Timing Explorer. Here, we apply a previously developed neutron star and accretion disc model to the NICER (Neutron star Interior Composition Explorer) data of the 2019 and 2022 outbursts of SAX J1808.4$-$3658. We find that a single circular hotspot model is insufficient to explain the data. Modelling with two hotspots and an accretion disc model provides better phase-residuals, but a spectral residual at around 1~keV remains. In contrast, we find a good fit with a flexible background approach, replacing the accretion disk. However, the inferred parameters are not robust due to a degeneracy in the origin of the non-pulsed radiation, which can be caused either by the background or a hotspot that is at least partially in view throughout a full rotation. This work represents an important next step in pulse profile modelling of AMPs by analysing NICER data and underlines the need for more accurate accretion disc and hotspot modelling to achieve robust parameter constraints. We expect the inclusion of higher energy and polarimetric data will provide complementary constraints on inclination, hotspot colatitude, and hotspot size, improving the accuracy of pulse profile modelling of AMPs.

\end{abstract}



\section{Introduction}
\ac{PPM} is a relativistic ray-tracing technique that models phase and spectrally-resolved pulse profiles from X-ray pulsars. It is mainly used to infer model parameters from the pulse profiles of millisecond pulsars, especially the mass and radius \citep{Watts2019b}. Measuring mass and radius provides a macroscopic probe of the \ac{EoS} of dense matter inside \ac{NS} cores (for reviews, see e.g. \citealt{Lattimer2016, Drischler2021}). The geometric parameters such as spin axis inclination and parameters of the X-ray emitting hotspots are also constrained, and these probe the magnetic field of the \acp{NS} as well as the surface. 

PPM is applied extensively to \ac{NICER} data of \acp{RMP}. \ac{NICER}, launched in 2017, is an excellent timing and spectroscopy instrument for the analysis of pulse profiles of \acp{RMP}, sensitive in the soft X-ray (0.2--12 keV). In \acp{RMP}, the hotspots are the magnetic polar caps, heated by bombardment of particles produced by pair creation in the \ac{NS} magnetosphere. To date, \ac{PPM} with \ac{NICER} data has provided both masses and radii and mapped the surface magnetic field of four \acp{RMP}: PSR J0740+6620 \citep{Miller2021, Riley2021, Salmi2022, Salmi2024a,Hoogkamer25}, PSR J0030+0451 \citep{Miller2019, Riley2019, Vinciguerra2024}, PSR~J0437$-$4715 \citep{Choudhury2024b}, and recently PSR~J0614$-$3329 \citep{Mauviard2025}. A fifth pulsar, PSR~J1231$-$1411, has also been analysed, but results are not fully conclusive \citep{Salmi2024b, Qi2025}.

This technique has also been applied to \acp{AMP}. Notably, \citet{Salmi2018}, hereafter \citetalias{Salmi2018}, used archival time- and energy-resolved \ac{RXTE} data to constrain parameters of the canonical \ac{AMP}, \aclu{J1808} (hereafter J1808), building on earlier analyses by e.g. \citet{Poutanen2003} and \citet{Kajava2011}. The \ac{PPM} technique can also be applied to the \acp{TBO} that are sometimes produced by accreting neutron stars \citep[e.g.][]{Watts2012, Kini2024b}.

In \acp{AMP}, material is accreted from the inner region of an accretion disc onto the \ac{NS} surface during an outburst. The charged particles in the accretion stream are funnelled towards the magnetic poles on the surface of the rotating \ac{NS}, giving rise to X-ray pulsations. These events are more luminous, with an x-ray luminosity $L_x\sim10^{36}-10^{37}$ erg/s \citep{Patruno2021}, compared to \acp{RMP}, with $L_x\sim10^{30}-10^{32}$ erg/s \citep{Becker1997Xray}, so shorter exposure times yield sufficient photons with \ac{NICER} for pulse profile analysis. Due to the presence of high energy electrons near the surface, inverse Compton scattering of seed photons takes place, giving rise to a power-law-like spectrum with a cut-off around 100~keV \citep{DiSalvo2022}. Unlike in \acp{RMP}, the electron scattering also causes the radiation to be significantly polarized providing an independent observable that constrains observer inclination and hot spot colatitude \citep{Viironen2004}.

Emission models for \ac{PPM} of \ac{AMP} have been becoming more accurate over time. Due to high computational cost, the usage of approximate analytical formulae for anisotropy and empirical models for Comptonized spectra in non-polarized emission models has been common \citep[e.g.][]{Poutanen2003, Salmi2018}. For polarized emission, a model was developed by \citet{Viironen2004} based on Compton scattering in an optically thin atmosphere but uses the Thomson scattering approximation. A description for a spherical star was derived by \citet{Poutanen2020} and for the oblate Schwarzschild approximation by \cite{Loktev2020}, and was recently applied to simulated data by \citet{Salmi2021}. More recently, \citet{Bobrikova2023} provided a model with higher accuracy that assumes Compton scattering in a slab geometry of hot electrons. Compared to sophisticated self-consistent accretion heated atmosphere models such as those developed by \citet{Suleimanov2018}, this model has fewer free parameters and is therefore well suited for \ac{PPM}.

\acp{AMP} are more challenging to constrain compared to \acp{RMP}. Firstly, this is because of additional elements that need to be accounted for in the model, such as the accretion funnel \citep{Ahlberg2024} and accretion disc. These introduce model parameters which will need to be explored during the \ac{PPM} analysis, making the process more computationally expensive. Secondly, the pulses of \acp{AMP} can also shift over time, because they depend on the variable accretion rate. This can sometimes pose a challenge, because an extended period of stable pulsations is required to gather enough photons in a pulse profile. A bias could be introduced if parameters are in reality shifting while they are assumed constant for a given pulse profile (see also \citealt{Kini2023}). However, the polarized X-ray pulsations they may provide could also boost constraints \citep{Viironen2004, Salmi2025}, and recently polarized radiation was indeed detected with a polarization degree at 4 percent from the \ac{AMP} SRGA J144459.2-604207 \citep{Papitto2025}.

Despite the challenges, their study is worthwhile because \acp{AMP} are an interesting group of sources. First of all, being potential progenitors of \acp{RMP}, they are interesting from a NS evolution perspective \citep{Alpar1982, Bhattacharya1991, Manchester2017}. Secondly, they are interesting from a mass-radius inference perspective, given that they exhibit multiple accretion related phenomena (e.g. X-ray bursts and persistent pulsations) and thus independent techniques can be applied for the same stars  \citep[e.g.][]{Salmi2018, Bult2019, Goodwin2019}.

Recently, simulations done by \citet{Dorsman2025}, hereafter \citetalias{Dorsman2025}, adapted the existing \ac{PPM} pipeline for \acp{RMP} and established that parameter recovery is possible for simulated \ac{NICER} \ac{AMP} data. They recovered parameters with tight 68 per cent \acp{CI}: $\pm 7$ per cent on mass $M$ and $\pm 6$ per cent on equatorial radius $R_{\rm eq}$ for one simulated \ac{AMP} scenario (A) with a large hotspot. For another scenario (B), where the hotspot was smaller and the \ac{AMP} was viewed more edge-on, they found slightly larger \acp{CI} and a slight bias in $M$ with the true value being outside the 68 per cent \ac{CI}. \citet{Salmi2025} performed a similar analysis on simulated polarimetric \ac{IXPE} data and while they found no good constraints on scenario A, they found good constraints for the inclination and hotspot colatitude on scenario B and other scenarios they considered that featured detectable polarized pulses.

In this paper we apply the analysis pipeline set up by \citetalias{Dorsman2025} to the canonical \ac{AMP} \ac{J1808}. More specifically, we will analyse the \ac{NICER} data of the most recent and only two outbursts observed by \ac{NICER}, in 2019 and 2022. Timing analyses of these data have been done by \citet{Bult2020} and \citet{Illiano2023}, respectively.

\ac{J1808} was the first \ac{AMP} to be discovered \citep{Wijnands1998}, has since gone into outburst 8 times, and has been extensively observed with X-ray (timing) telescopes such as \ac{RXTE} and \textit{XMM-Newton} (for a review, see e.g. \citealt{Patruno2021}). Early studies of the spectrum and phase-resolved persistent pulsations were able to fit \ac{RXTE} data well with one or two blackbody components and a Comptonization component \citep{Heindl1998, Gierlinski2002, Kajava2011}. \citet{Gierlinski2002} and \citet{Poutanen2003} also included a Compton reflection component ($>$10~keV) with a broadened iron line (6--7 keV). The iron line is also found with other high-energy instruments and has been the subject of in-depth study \citep[e.g.][]{Wilkinson2011, DiSalvo2019, Sharma2023}. The broadening of the iron line has been used to estimate the inner disc radius \citep{Cackett2009, Papitto2010}. Besides persistent pulsations, \ac{J1808} also exhibits thermonuclear bursts and \acp{TBO} (\citealt{Chakrabarty2003, Bult2019}, for a review, see \citealt{Bhattacharyya2022}). X-ray bursts have also been used to estimate the distance to \ac{J1808} \citep[e.g.][]{Galloway2006, Galloway2024}. 

This work addresses some gaps in previous \ac{PPM} analyses of \acp{AMP}. To start, we apply \ac{PPM} for the first time to new \ac{NICER} data of an \ac{AMP}, extending the \ac{PPM} of \acp{AMP} to newer outbursts. This is a lower energy band than e.g. \ac{RXTE}, and is thus complementary to higher energy data, being sensitive to both the lower energy end of the hotspot radiation and the disc blackbody radiation. However, these components overlap and are therefore challenging to constrain individually \citepalias{Dorsman2025}. Additionally, data in this band is complementary, but not uniquely positioned to constrain the disc, because past high energy data has also been used to constrain the disc through its other signatures, such as light path obscuration \citep{Kajava2011}, reflection \citep{Ibragimov2009} and iron line broadening \citep{Cackett2009, Papitto2009}, neither of which are applied here. We employ also a more recent atmosphere model from \citet{Bobrikova2023} compared to \citetalias{Salmi2018}, who use an empirical model for the Comptonization spectrum and an analytical parametrization for the angular dependence of the radiation. Furthermore, we make use of the pipelines set up for \ac{NICER} \acp{RMP} in the open source \ac{X-PSI} software package, making this work very easily reproducible. Additionally, the model in our analysis also includes two hotspots, compared to one in \citetalias{Salmi2018} and \citetalias{Dorsman2025}. Lastly, here we also explore the usage of a multicolour accretion disc model and, alternatively to the disc model, the usage of a marginalised background approach that accounts for non-pulsed counts in each energy bin independently (see \Cref{sec:model} for more details).

This paper is structured as follows. \Cref{sec:method} describes the methodology, including the methodology for parameter inference and the model. \Cref{sec:data} describes the preparation of the \ac{NICER} data. \Cref{sec:results} gives the results, while \Cref{sec:discussion} discusses the results and puts them into context. Finally we conclude in \Cref{sec:conclusion}.

\section{Methodology}\label{sec:method}
\subsection{Parameter inference and likelihoods}\label{sec:bayesian_inference}
The main goal of this work is to estimate parameters of the \ac{AMP} \ac{J1808}. The method we employ is \ac{PPM}, in which we fit a millisecond pulsar model, an \ac{AMP} in this case, to a measured pulse profile. Much of the methodology is identical to \citetalias{Dorsman2025}, so in this section we only give a brief overview of the methodology that stays the same while referring back to that work for more in-depth explanations. However, we detail methodological changes and additions here.

\ac{PPM} is an instance of Bayesian parameter inference, and in this work, we use it to estimate \acp{PPD} of parameters. As a reminder, the posterior distribution $P(\theta|D,M)$ of a set of parameters $\theta$ conditional on a dataset $D$ and model $M(\theta)$ can be expressed with Bayes' theorem:

\begin{equation}\label{eq:posterior}
P(\theta|D,M)=\frac{P(D|\theta,M)\pi(\theta|M)}{P(D|M)}.
\end{equation}
Here, $P(D|\theta,M)=L(\theta)$ is the likelihood function, $\pi(\theta|M)$ is the prior distribution and $P(D|M)= Z$ is the evidence or marginal likelihood. The evidence is independent of model parameters $\theta$, so it plays the role of a normalisation factor for the \ac{PPD}. Additionally, the ratio of the evidences for two different models, also called the `Bayes factor' is commonly used for the purpose of model comparison.

In practice, we estimate the \acp{PPD} and evidences by Nested sampling \citep{Skilling2004} with \textsc{MultiNest} \citep{Feroz2009, Feroz2019}. Identically to \citetalias{Dorsman2025}, we set the sampling efficiency in \ac{X-PSI} to 0.1 and use 1000 live points (unless otherwise stated).

Compared to \citetalias{Dorsman2025}, who use only a Poissonian likelihood function, we also use a Poissonian likelihood function with marginalised background in some cases. The latter likelihood function is given by
\begin{equation}\label{eq:marginalisatoin}
    L_{\rm marg}(\theta) := \int L(D|\theta, M, \bm{B})\pi(\bm{B})d\bm{B}, 
\end{equation}
where $\pi(\bm{B})$ is the prior on the background and $L(D|\theta, M, \bm{B})$ is the Poissonian likelihood function. Each element of the vector $\bm{B}$ is some phase-constant number of background counts $B_i$ for each energy channel $i$, which are added to the \ac{NS} counts. For a given pulse profile, all options for $B_i$ that are allowed by the prior contribute to the integral. However, the best (highest likelihood) pulse-profile plus background combinations contribute the most towards the likelihood integral. More detail on this topic can be found in section 3.2.10 and Appendix B of \cite{Riley2019thesis}.

Marginalising the background was not used in \citetalias{Dorsman2025}, because the expected instrumental and astrophysical backgrounds were expected to be small compared to the bright \ac{AMP}. Additionally, they included the source background (i.e. phase-constant contribution from the source) explicitly in the model: blackbody radiation from the accretion disc. They assumed that component would dominate in the \ac{NICER} band. While that approach has the benefit of adhering solely to physically motivated radiation components, it is also inflexible if some radiation components are not correctly accounted for in the modelling. In this work we also include background marginalisation as an alternative, more flexible, approach.

\subsection{Model}\label{sec:model}
For the (pulsed) radiation from the \ac{NS} we use \ac{X-PSI} \citep{Riley2023}. \ac{X-PSI} is a code for the forward modelling of time-dependent X-ray data from a pulsar, as well as for Bayesian inference of model parameters. We use \ac{X-PSI} resolution settings throughout this work that are the same as in table 1 of \citetalias{Dorsman2025}.

The equations that govern the time-dependent radiation from surface anisotropies of \acp{NS} (and used in \ac{X-PSI}) has been well-developed in a body of work that spans decades \citep[see e.g.][]{Pechenick1983, Riffert1988, Miller1998, Poutanen2003, Bogdanov2019b}. Important physical effects are included such as the oblate \ac{NS} shape + Schwarzschild approximation \citep{Morsink2007, AlGendy2014}, gravitational redshift, light bending, Doppler boosting, and time delay due to light path difference. For \acp{AMP} specifically, we use the Compton slab atmosphere derived by \citet{Bobrikova2023} for the hotspot atmospheres, and the blackbody disc model (discussed below) as the soft X-ray contribution of an accretion disc. Finally, the X-ray radiation to be received by the observer is first attenuated by the neutral hydrogen column in the interstellar medium and then convolved with the response of \ac{NICER}. \Cref{tab:parameters} lists all the parameters along with their descriptions. 

\begin{table*}
    \caption{\label{tab:parameters} Model parameters and prior distributions. $U$(lower, upper) is a uniform distribution with lower and upper bounds and $N(\mu,\sigma)$ is a normal distribution with mean $\mu$ and a standard deviation $\sigma$. For normal distributions the cut-off is at 5$\sigma$. The subscript ${\rm p}/{\rm s}$ indicates these parameters exist for both primary and secondary hotspots, where secondary is only present for \stu\ models (see the text in \Cref{sec:model}).}
    \begin{tabular}{lll} \hline \hline
    Parameter (Unit) & Description & Prior Density  \\ \hline
    \multicolumn{3}{c}{Pulsar}\\
    $D$ (kpc) & Distance & $N(2.7,0.3)$ \\
    $M$ (M$_\odot$) & Mass & $U(1,3)^a$ or EoS informed$^b$  \\
    $R_\mathrm{eq}$ (km) & Equatorial radius & $U(3R_{\rm G}(1),16)^a$ or EoS informed$^{b}$ \\
    $\cos i$ (-) & Cosine inclination & $U(0.15, 0.87)$\\
    $f$ (Hz) & Pulsar frequency & fixed at 401\\
    $N_\mathrm{H}$ (\SI{e21}{cm^{-2}}) & ISM column density & $N(1.17,0.2)$\\\hline
    \multicolumn{3}{c}{Hotspots}\\
    $\phi_{\rm p/s}$ (cycles) & Phase & $U(-0.25,0.75)$ \\
    $\cos\theta_{\rm p/s}$ & Cosine co-latitude & $U(1, -1)$ \\
    $\zeta_{\rm p/s}$ (deg) & Angular radius & $U(0, 90)$\\
    $T_{\rm seed,p/s}$ (keV) & Seed photon temperature & $U(0.5,1.5)$ \\
    $T_{\rm e,p/s}$ (keV) & Electron slab temperature & $U(20,100)$\\
    $\tau_{\rm p/s}$ (-) & Thomson optical depth & $U(0.5,3.5)$\\\hline
    \multicolumn{3}{c}{Disc}\\
    $T_{\rm in} (\log_{10}\rm K)$ & Inner disc temperature & $U(5.06,6.84)^c$\\
    $R_{\rm in}$ (km) & Inner disc radius & $U(R_{\rm eq},R_{\rm co})^d$\\\hline
    \multicolumn{3}{c}{Gaussian Line}\\
    $\mu$ (keV) & Mean & $U(0.8, 1.1)$\\
    $\sigma$ (keV) & Standard deviation  & $U(0.01, 0.5)$\\
    $N$ (10$^{37}$ photons/cm$^2$/s) & Normalisation & $U(0.01, 10)$\\\hline
    \end{tabular}\\ 
    \vspace{2 mm}
    \begin{flushleft}
    \footnotesize{$^a$ These priors are also bound by the causality limit for compactness. $R_{\rm G}(1)$ is the gravitational radius of $M=1$M$_\odot$. More detail is given in the text in \Cref{sec:prior}.}\\
    \footnotesize{$^b$ In runs that use the \ac{EoS} informed prior, this prior is replaced. More detail is given in the text in \Cref{sec:prior}.}\\
    \footnotesize{$^c$These limits correspond to (0.01, 0.6) in keV. In \citetalias{Dorsman2025}, $T_{\rm in}$ was also uniform in $\log_{10}\rm K$, despite being incorrectly listed as uniform in keV.}\\
    \footnotesize{$^d$ $R_{\rm co}$ is a function of $M$ and $f$. See \Cref{sec:prior}.}
    \end{flushleft}
\end{table*}

Regarding the hotspots on the \ac{NS}, \citetalias{Dorsman2025} used only a single circular hotspot model. Following the nomenclature introduced in \citet{Riley2019}, we call this hotspot configuration \st. Here, `single' refers to the uniform temperature profile on the hotspot. From \ac{MHD} simulations it is not expected that the radiation profiles on hotspots should be uniform \citep[e.g.][]{Romanova2004, Das2025}, but computational effort is significantly reduced under this simplification, and even more so with the computational optimization achieved in \citetalias{Dorsman2025}.

In this work, we will use also two circular hotspots, referred to as \stu, where U is `unshared', which refers to the two hotspots having unshared parameters between each other. The two hotspots are not allowed to overlap and the hotspot with the smaller colatitude is always referred to as the `primary' with the other as the `secondary'. 

Regarding the non-pulsed accretion disc, \citetalias{Dorsman2025} used \texttt{diskbb}, a multicolour disc blackbody \citep{Mitsuda1984, Makishima1986}. No gravitational redshift or spectral hardening is included in this model. In addition, no interaction between the radiation of the disc and the star is implemented, such as obscuration or reflection by the disc of the radiation from the hotspot. 

Here, we introduce a model component to the disc which is a simplified model to imitate a broadened (reflection) line. This is motivated by some evidence for a broadened line contribution around 1 keV in the residual found in the analysis with only an accretion disc background detailed in \Cref{sec:results}. To model the broadened line we use a Gaussian distribution for the flux per energy
\begin{equation}\label{eq:line}
    F_{\rm line}(E)=\frac{N}{\sigma\sqrt{2\pi}} e^{-\frac{(E-\mu)^2}{2\sigma^2}},
\end{equation}
where $E$ is the photon energy, $N$ is the normalisation, $\mu$ is the mean and $\sigma$ is the standard deviation. 

In addition to modelling the source background with a blackbody disc or blackbody disc with a broadened spectral line, we also include a marginalised background as a more flexible alternative. In \ac{RMP} analyses, this feature has been used to capture non-source background counts. However, in this case we use this feature to replace the disc model, and thus account for non-pulsed background counts that originate from the accreting environment around the star. The non-source counts, although peaking at an estimated 2 per cent of counts, are ignored here because they average out to much less than 1 per cent of counts in the data.

To use the marginalised background, an upper and lower bound must be imposed for each energy bin. The widest possible bounds are zero counts for the lower bound and background counts matching the data for the upper bound. In preliminary testing, we found that setting these bounds led to biased background marginalisation. Specifically, the inferred background count rate would be biased to be significantly higher if the true background count rate is near zero. To evade this situation, we constrain the background further by using a fiducial background flux $f_i$ for each bin $i$ and multiplicative `support' factor $s$ around it: [lower$_i$, upper$_i$] = $[f_i/s, f_i*s]$. 

Expecting the background to be dominated by the blackbody radiation from an accretion disk, we use a previously obtained fit of the 2019 data with \texttt{diskbb} for $f$. As discussed by \citetalias{Dorsman2025}, it is difficult to disentangle the disc and star contributions with only \ac{NICER} data, so it is difficult to make a robust estimate for the fiducial disc. Nevertheless, we use the disc flux from scenario A in \citetalias{Dorsman2025}, which is a fit of the 2019 \ac{NICER} pulse profile of \ac{J1808}. This disc flux was found by fitting using a one hotspot model, and by fixing parameters $M$ (mass), $R_{\rm eq}$ (equatorial radius), $D$ (distance), and $N_\mathrm{H}$ (neutral hydrogen column). We choose an arbitrary large $s$, initially at 100, with the aim of providing a sufficiently large deviation space around the fiducial disc flux, and therefore reducing the dependence on the initial choice for fiducial disc. However, we found the results still depend on this choice and we investigate this further in \Cref{sec:stum}. 

\subsection{Prior density distributions}\label{sec:prior}
Another required ingredient of Bayesian analysis are prior probability distributions, which encode a priori known information about the model parameters, i.e. before considering the data. \Cref{tab:parameters} gives an overview of the priors in the right-most column. The priors are mostly identical to those used in \citetalias{Dorsman2025}. In this section we describe the priors with a focus on changes compared to that paper. For some parameters we also give a more detailed account of estimates in the literature and describe how those influenced the choices of priors here.

Distance estimates of \ac{J1808} have been in the range 2--4 kpc. Recently, matching 2019 data of Type-I X-ray bursts, \citet{Goodwin2019} estimated 3.3$^{+0.3}_{-0.2}$ kpc. This estimate comes from a comparison of the observed bursts with a theoretical ignition model which also takes into account other parameters including anisotropy, fuel composition, and 
\ac{NS} mass and radius. More recently, \citet{Galloway2024} reanalysed the same data with updated code and estimated a smaller distance of $2.7\pm0.3$ kpc, where this reduction appeared to be driven by higher burst anisotropy. We use a normal distribution for the distance corresponding to this latest estimate.

The hydrogen column density parameter $N_{\rm H}$ has a large effect in the lower end of the \ac{NICER} energy band (below ~1 keV). Estimates for $N_\mathrm{H}$ in the direction of \ac{J1808} have been done in the past. 
\citet{Patruno2009a} found $(1.4 \pm 0.2)\times10^{21}$cm$^{-2}$ deriving the equivalent hydrogen column density by measuring the spectrum of \ac{J1808} from absorption lines around the oxygen K absorption edge. \citet{Papitto2009} found a larger value of $2.14^{+0.02}_{-0.03} \times10^{21}$cm$^{-2}$ when fitting the continuum spectrum. We use $1.17\times10^{21}$cm$^{-2}$ as the mean value of a normally distributed prior for $N_{\rm H}$. This value was computed with the HEASoft nH tool \citep{HEASoft} and the HI4PI map \citep{bekhti2016}, which is based on 21-cm radio observations and reports values of the neutral hydrogen column to the edge of the galaxy. Because there is uncertainty in this value we use a large $\sigma=0.2\times10^{21}$cm$^{-2}$ with a cut-off at $5\sigma$ so that this distribution at least covers the values found in the literature. Note that in our model $N_{\rm H}$ scales with the extinction contribution, but the energy dependence of the extinction is set by the relative abundances for the interstellar medium from \cite{Wilms2000}.

As in \citetalias{Dorsman2025}, we use flat and wide priors for the mass $M$. The equatorial radius $R_{\rm eq}$ has a fixed lower limit at $3 R_G(1)=4.4$ km and upper limit at 16 km, where $R_{\rm G}(1)$ is the gravitational radius of $M=1$M$_\odot$. These two priors are modified by rejection of samples that have too high compactness: beyond the causality limit given by $R_{\rm pole}/R_{\rm G}(M)>2.9$ \citep[see e.g.][]{Gandolfi2012}, where $R_{\rm pole}$ is the polar radius. These $M$ and $R_{\rm eq}$ priors are very wide and cover current dense matter models (see also \citealt{Riley2021}). Results obtained here with these priors can in principle be used to constrain \ac{EoS} models, while remaining independent of any a priori assumptions on the \ac{EoS} model, results from other pulsars, \acp{GW}, and nuclear experiments \citep{Riley2018}.

Besides the previously used priors, we will also use an alternative set of priors for $M$ and $R_{\rm eq}$ conditional on a choice of \ac{EoS} model and measurements from other studies. Because of the prior conditionality, \acp{PPD} inferred using that prior will likely be consistent with studies it is conditional on, so cannot be used to verify or test those other studies. Also due to the conditionality, these \acp{PPD} would be disqualified from usage in deriving further \ac{EoS} constraints, at least without accounting for the fact that they are already conditional on an underlying \ac{EoS} model choice and related parameter choices.

A major motivation to use such an \ac{EoS}-informed prior anyway, is that it would likely save significantly on computational time to explore a smaller prior space. While the results will not be independent from other measurements, that is perhaps justified (at least in a Bayesian sense) given that we expect that the \ac{EoS} is universal among \acp{NS}, and the now large number of mass-radius inferences from other works. For now, our goal is merely exploratory, i.e. to test whether the informed prior leads to comparable evidences, and how inferred parameters would be affected.

We use as our \ac{EoS}-informed mass-radius prior the recently constrained dense matter \ac{EoS} by \citet{Rutherford2024}. They derive a mass-radius \ac{PPD}, which we use as a prior, through a Bayesian analysis of \ac{NICER} pulsar data and tidal deformability measurements via \acp{GW}. They analyse these data with various options as their \ac{EoS} model and priors. We choose their ``New'' scenario, which corresponds to the green posterior in the bottom-left panel of their Fig.\,5. This scenario includes recent N$^3$LO $\chi$EFT calculations of particle interactions in dense matter by \citet{Keller2023} up to a transition density and onwards parametrises the \ac{EoS} with a piecewise polytropic (PP) model. We use their results with a transition density at 1.5$n_0$, where $n_0$ is the saturation density. 

We apply their results by fitting a one-dimensional CDF for both $M$ and $R_{\rm eq}$ using all the samples. This is a similar approach to that used by \citet{Salmi2024b} for $R_{\rm eq}$. This simplification was necessary because jointly drawing a sample from a two-dimensional prior is not straightforwardly possible in \ac{X-PSI}. Nevertheless, this is an acceptable simplification for our purpose, which is merely exploratory. When using this approach for headline results, a two-dimensional fit would be recommended.

There have been a number of studies that place constraints on the inclination $i$ of the system: To start, $i<82^{\circ}$ was derived by \citet{Chakrabarty1998} from the absence of X-ray eclipses. \citet{Poutanen2003} set a limit $i>65^{\circ}$ from modelling of pulse profiles. \citet{Deloye2008} derive $i=36-67^{\circ}$ based on optical observations throughout the binary orbit, which lead to constraints on the two masses and inclination. \citet{Cackett2009} find $i=51-63^{\circ}$ at 90 per cent confidence, from joint fitting of the X-ray broadband spectrum and Fe K$\alpha$ iron-line. \citet{Ibragimov2009} constrain the inclination to $i=50-70^{\circ}$ based on modelling of the pulse profiles of the 2002 outburst, while \citet{Kajava2011} analyse the pulse profiles of the 2008 outburst and estimate the inclination to be $i=58^{\circ}{_{-6}^{+4}}$. \citet{Morsink2011} analyse pulse profiles of multiple epochs and set a limit of $i>41^{\circ}$. \citep{DiSalvo2019} find that $i>50^{\circ}$ based on the broad band spectrum and iron line. Finally, \citet{Goodwin2019} find $i=69^{\circ}{^{+4}_{-2}}$ based on the analysis of thermonuclear X-ray bursts.

We set as upper limit $i\lesssim81^{\circ}, \cos{i}>0.15$ \citep{Chakrabarty1998}. As a lower limit we use $i\gtrsim30^{\circ},\cos{i}<0.87$, which leaves some extra room below the lower limits found by previous studies. We use a uniform prior between these boundaries in $\cos{i}$ space to uniformly sample the viewing angle on a sphere.

The parametrisation of the hotspot geometry is simplified to circles, meaning their position and shape are governed by the phase $\phi$, colatitude (or magnetic obliquity) $\theta$, and angular radius $\zeta$. We use uniform priors that cover the full parameter space for these parameters. To sample uniformly from a sphere, we adopt a uniform prior on the cosine of the colatitude. New in this work compared to \citetalias{Dorsman2025} is that we will allow two hotspots. These hotspots are mutually independent and therefore not necessarily antipodal. However, as described in \Cref{sec:model}, the hotspots are ordered and are not allowed to overlap. In practice (during sampling) hotspots that do not adhere to this rule are rejected before costly model computation.

Similarly we adopt uniform priors for the parameters that govern the hotspot atmosphere: electron temperature $T_{\rm e}$, seed photon temperature $T_{\rm seed}$, and optical depth $\tau$, where the bounds are such that the full precomputed dataset by \citet{Bobrikova2023} is utilised.

For the inner disc temperature $T_{\rm in}$ we use a broad prior, uniform in $\log_{10}T$, with $T$ in Kelvin. The lower boundary is at 5.06 (0.01 keV) and the upper boundary at 6.84 (0.6 keV). This covers a range of values that have been inferred in other works, such as 0.2 keV \citep{Patruno2009a} and 0.3 keV \citep{Kajava2011}. We note that higher values for $T_{\rm in}$ are inferred in spectral analysis in other works that analyse the 2019 and 2022 \ac{NICER} data, and contemporaneous \texttt{AstroSat} data \citep{Bult2019,Sharma2023,Kaushik2025}. However, in these cases, the seed temperature of the Comptonisation component are tied to the $T_{\rm in}$ of the \texttt{diskbb} model, meaning that the radiation from the disk is Comptonised by a surrounding corona. This astrophysical assumption, coupled with the absence of a \ac{NS} hotspot component, leads to a higher $T_{\rm in}$, around 0.5 to 1 keV. In contrast, in this work we assume that the radiation from the hotspots is Comptonised at the surface and that there is no corona.

As a prior for the inner disc radius $R_{\rm in}$, the value is uniformly drawn between the $R_{\rm eq}$ and co-rotation radius $R_{\rm co}=(GM/4f^2\pi^2)^{1/3}$, where $G$ is the gravitational constant. Note here that $M$ is another parameter in the model, so the prior will cut off at a different value for each sample. When making many draws from the prior, the shape looks flat and then slopes down at higher values of $R_{\rm in}$. $R_{\rm co}$ is used as an upper limit because outside this radius the accretion would be in the propeller regime (see e.g. section 4.1.2 of \citealt{DiSalvo2022}), which is not expected during the peak of the outburst. If the system is in the (weak) propeller regime it would be worth exploring higher upper limits: simulations done by \citet{Romanova2018} have shown that accretion can proceed when $R_{\rm in}$ exceeds $R_{\rm co}$ by a factor of a few. If the inferred $R_{\rm in}$ is close to the upper limit, it could be prudent to explore alternative upper limits as well: \citetalias{Dorsman2025} found that the inferred mass was biased in their scenario B, in which the injected $R_{\rm in}$ value was near the upper limit (see their section 6.3 for further discussion).

\section{Data Preparation}\label{sec:data}
\subsection{NICER Observations} \label{sec:data_nicer}
In this work we analyse \ac{NICER} data from the 2019 and 2022 outbursts of \ac{J1808}. Analysing two datasets separately allows for a check of consistency for $M, R_{\rm eq}, D, N_\mathrm{H}, i$ between datasets, while other (hotspot) parameters may vary between outbursts. This section describes the preparation of both datasets.

\ac{NICER} is an external payload on the International Space Station, which contains the X-ray Timing Instrument (XTI). The XTI houses an array of 56 (52 operational) pairs of coaligned X-ray concentrator optics and silicon drift detectors in focal plane modules (FPMs). The fast-timing capabilities in the 0.2--12.0~keV energy range, the unprecedented effective area in the soft X-rays ($\sim 1900{\rm\,cm^2}$ at 1~keV), the 100~ns time-tagging accuracy afforded by the onboard global positioning system receiver \citep{Gendreau2016,LaMarr2016,Prigozhin2016}, and the flexible scheduling capabilities, make NICER the best current instrument to track the evolution of the X-ray pulsations of \ac{J1808}. 

\begin{figure}
  \centering
  \begin{minipage}[b]{0.48\textwidth}
    \centering
    \includegraphics[width=\textwidth]{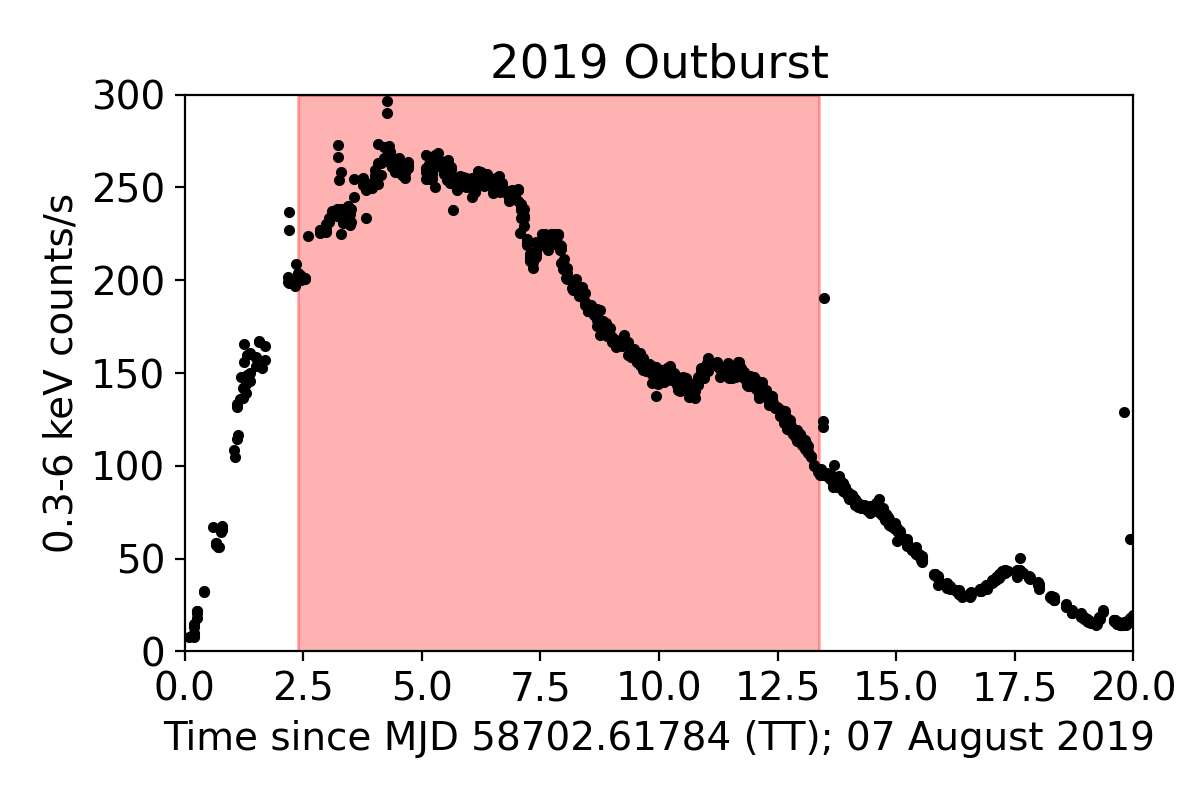}
    \label{fig:outburst2019}
  \end{minipage}
  \hfill
  \begin{minipage}[b]{0.48\textwidth}
    \centering
    \includegraphics[width=\textwidth]{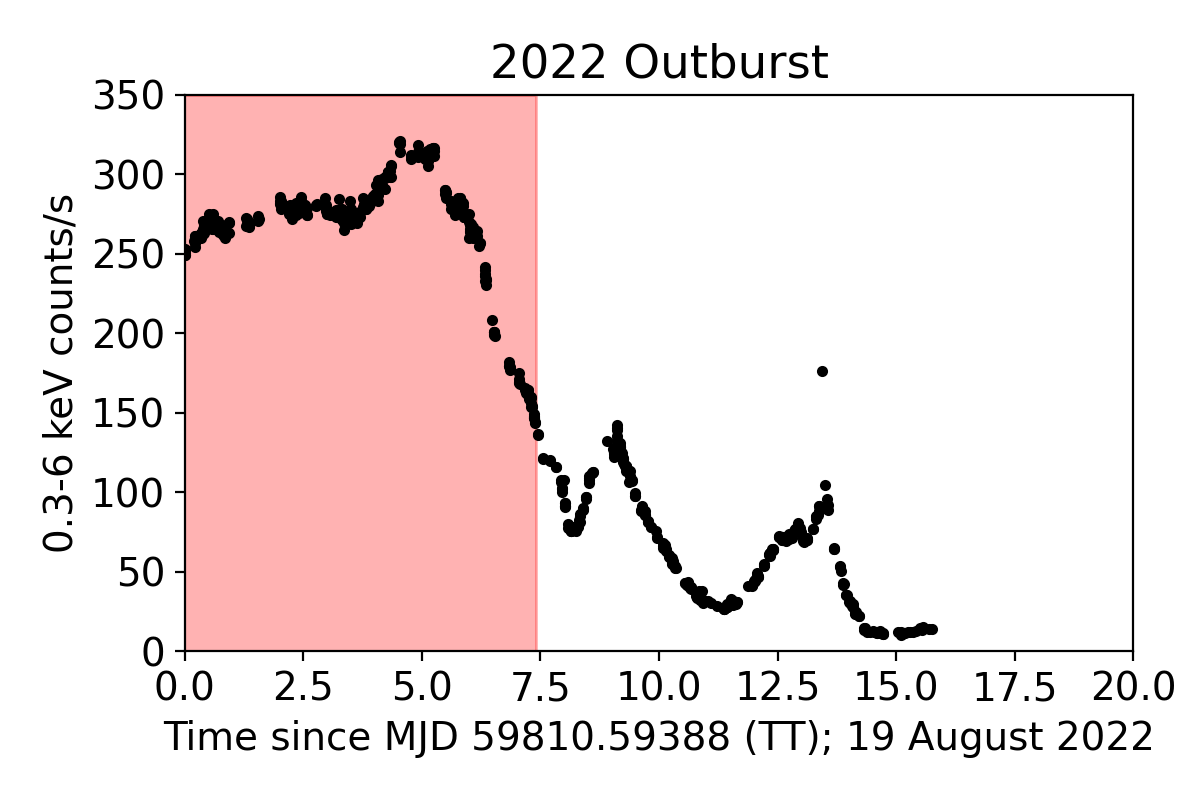}
    \label{fig:outburst2022}
  \end{minipage}
  \caption{The outbursts of \ac{J1808} in 2019 and 2022. The red band highlights the interval within which data was extracted for this analysis.}
  \label{fig:outburst}
\end{figure}

We conducted a preliminary analysis of the data where we ultimately decided to make use of observations during intervals where the brightness is around the peak. In the case of the 2019 outburst, as shown by \Cref{fig:outburst}, we used observations between 2019 August 10 and 2019 August 20 inclusive (a subset of `ObservationIDs' (ObsIDs) starting with 258401). For the 2022 outburst, observations between 2022 August 19 to 2022 August 26 inclusive (a subset of ObsIDs starting with 505026 and 557401) were utilized. All of the observations were reduced and processed with \textsc{HEASoft} v6.31.1 and the NICER Data Analysis Software (\textsc{NICERDAS}) v10a (2022-12-16\_V10a) using the NICER calibration database version \texttt{xti20221001}. In constructing the good time intervals (GTIs) for scientific analysis, we imposed the following filtering criteria: Earth limb elevation angle ${\rm ELV} > 15^\circ$; bright Earth limb angle ${\rm BR\_EARTH} > 30^\circ$; undershoot rate (per FPM; dark current) of ${\rm underonly\_range} = 0-500{\rm\,c/s}$; overshoot rate (per FPM; charged particle saturation) of ${\rm overonly\_range} = 0-30{\rm\,c/s}$; NICER transiting outside of the South Atlantic Anomaly; angular pointing offset for the source of ${\rm ANG\_DIST<54\arcsec}$. These resulted in roughly 132.4~ks and 71.3~ks of filtered exposure for the 2019 and 2022 outbursts, respectively.

\subsection{Responses}
\label{sec:data_responses}
The instrument response converts source photons to counts detected by \ac{NICER} per energy bin. We first conducted a preliminary analysis to assess whether individual response files are required for each ObsID analysed, or whether a representative response file would suffice. We generated the ancillary response files (arf; encodes effective area information) and the response matrix files (rmf; encodes energy redistribution information)\footnote{\url{https://heasarc.gsfc.nasa.gov/docs/nicer/analysis_threads/arf-rmf/}} for each ObsID using the \textsc{nicerarf} and \textsc{nicerrmf} tasks (with detectors 14 and 34 disabled) as has been done in recent work for PSR~J0740$+$6620 \citep{Salmi2024a} and PSR~J0437$-$4715 \citep{Choudhury2024b}. For each of the 2019 and 2022 outbursts, we inspected the fractional differences in the effective area curves (i.e., arfs) between the ObsIDs. For the 2019 outburst, the largest fractional differences were on the order of 3 per cent; though looking further, we noticed that in one time interval within ObsID 2584010201, there were only 7 detectors turned on (all other intervals had 50). For consistency, we excised the time interval;\footnote{NICER mission elapsed time of 177613405 to 177618415 (seconds, TT units)} the fractional difference fell to $\sim0.2$ per cent. For the 2022 outburst, the differences were of the order $\sim1$ per cent. We also compared the fractional differences between the averaged 2019 and averaged 2022 effective area curves, which was about 4.5 per cent. 

We did a similar exercise for the rmfs, and we found that over 99.8 per cent of the energy channels do not show significant deviations from an averaged rmf. Thus given the large fractional differences in the averaged effective area curves between the 2019 and 2022 observations, we constructed exposure-weighted average effective area curves and rmfs for each individual outburst, using \textsc{ftaddarf} and \textsc{ftaddrmf}, which are available within \textsc{HEASoft}.

To save on computational effort we exclude high energies where the count-rate is low. We restrict the instrument response (and corresponding data) to channels 0 to 570 (lower edge 0.3~keV until upper edge 6.0~keV). For both datasets this restriction only excludes 1 per cent of the counts.

\subsection{Pulse 
profiles} \label{sec:data_pp}
Here we discuss the observational variability of the X-ray pulse profile to justify our data selection choices. 
For each of the 2019 and 2022 X-ray outbursts, we initially folded the X-ray photons using the quadratic phase timing model presented in \citet{Bult2020} and the linear phase model from \cite{Illiano2023}. We then constructed custom local timing solutions for the data intervals specified previously by fitting with PINT \citep{Luo2021}. 

While the flux changes substantially throughout the data selections, we find that shapes of the pulse profiles are mostly stable. For the 2019 outburst, the selected data cover the last part of the rise, the peak, and much of the decline. The phase residual is small with the maximum phase difference $\sim$0.05 cycles. Meanwhile, the fractional amplitude is in a range between 4.5--5.5 per cent for the first part of the data, but starting at August 18 (re-brightening during the decline) it rises to a maximum of 6.3 per cent. To study the pulse shape evolution, we also compute the maximum difference in normalised count rates across ObsIDs in each phase bin and find that the difference stays within a maximum of 3 per cent. For the selected data from the 2022 outburst, the data covers the peak of the outburst and much of the decline. The maximum phase difference is larger compared to the 2019 outburst throughout, at $\sim$0.1 cycles. The fractional amplitude meanwhile decreases from 4.8 to 3.7 per cent during the peak, but increases back up to a maximum of 6.1 per cent during the decline. The shapes of the pulse profiles are similarly stable, but the maximum difference in normalised count rates peaks a bit higher at 4 per cent. 

\begin{figure}
  \centering
  \begin{minipage}[b]{0.48\textwidth}
    \centering
    \includegraphics[width=\textwidth]{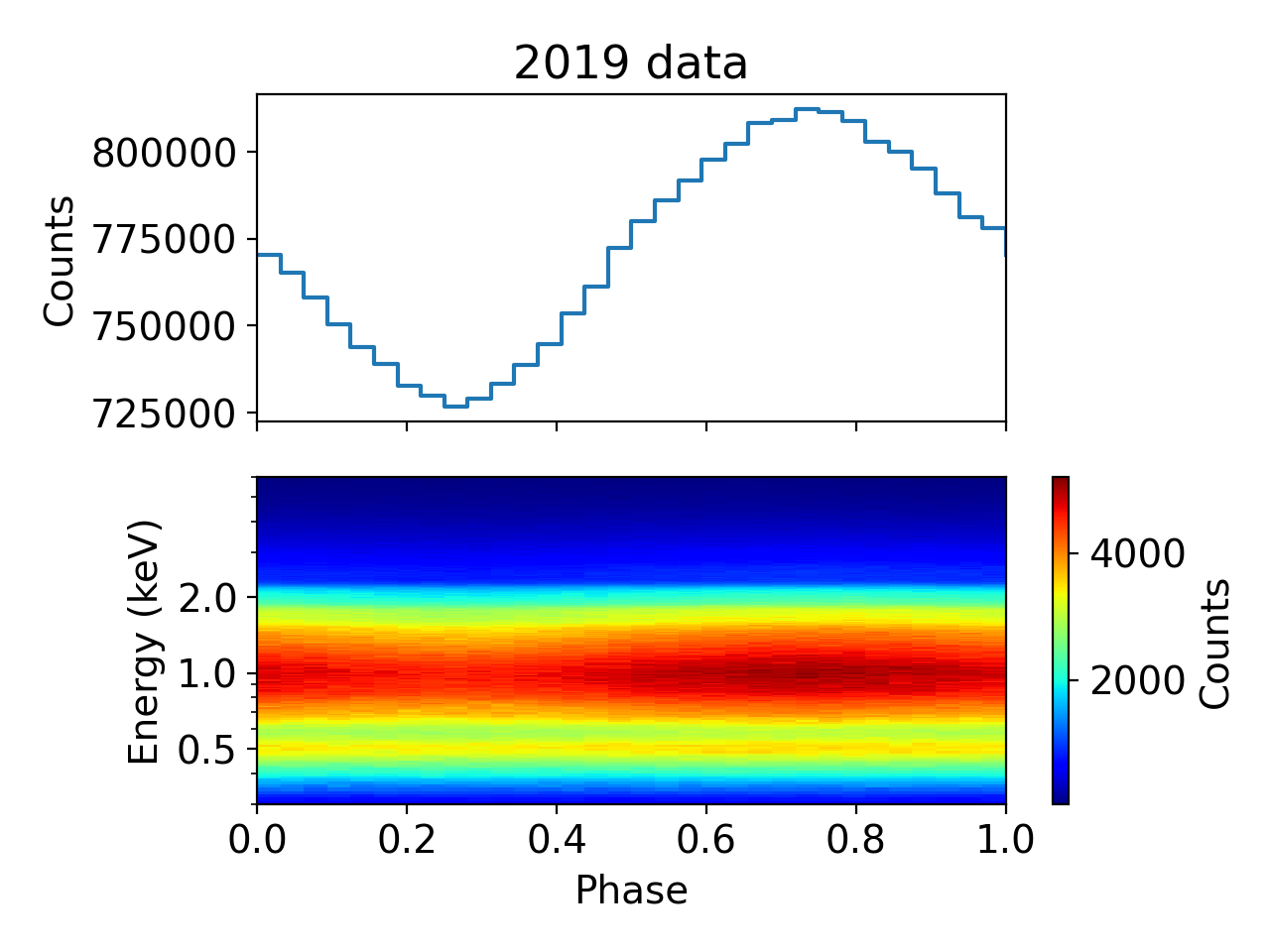}
    \label{fig:data2019}
  \end{minipage}
  \hfill
  \begin{minipage}[b]{0.48\textwidth}
    \centering
    \includegraphics[width=\textwidth]{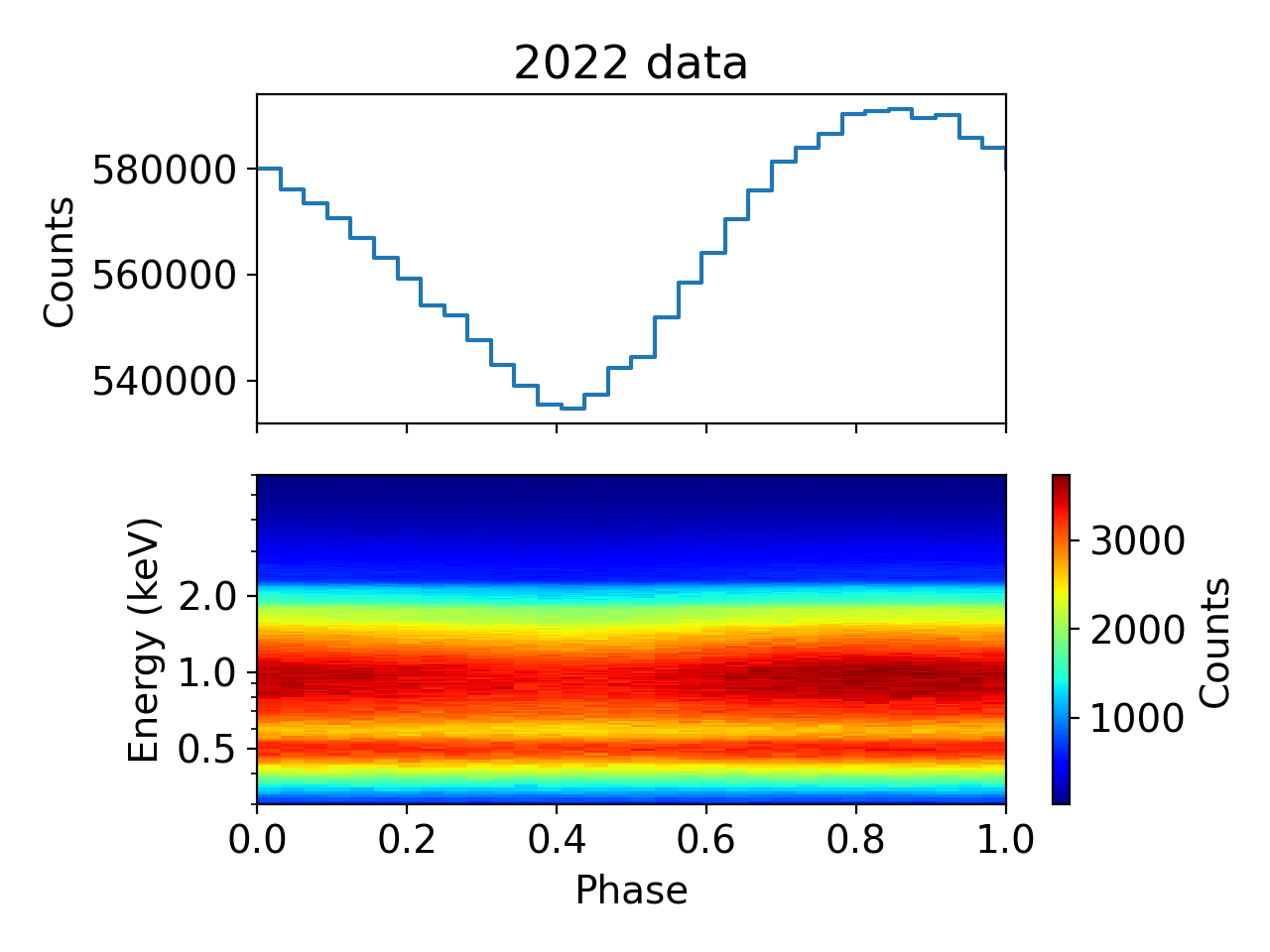}
    \label{fig:data2022}
  \end{minipage}
  \caption{Pulse profiles prepared from the peak of the 2019 and 2022 outbursts of \ac{J1808}. The top panels display the bolometric pulse profiles while the bottom panels display the energy-phase resolved pulse profiles.}
  \label{fig:data}
\end{figure}

The increase in pulse fraction correlates with a decline in flux and is likely caused by the reduction in accretion rate. There could thus be variability of model parameters related to accretion (e.g. $T_{\rm seed}, T_{\rm in}, R_{\rm in}$ and $\theta_{\rm p/s}$) which is averaged out within this data selection. The effect of ignoring parameter variability was studied within the context of \acp{TBO} by \citet{Kini2023}, and they found a bias in inference of $M$ and $R$. However, we recommend a separate study with synthetic data to quantify any biases this could cause within this context, where the change in flux is smaller. \Cref{fig:data} presents the pulse profiles that result from this data preparation.

\section{Results}\label{sec:results}
This section details the results of the analysis. \Cref{sec:singlehotspot} describes the results of the analysis done using a model with a single hotspot (\st). This section includes three different background treatment approaches: \disc, \discline, and background marginalisation. However, in all cases the best model fits left visible residual structures. As a result, we also do \ac{PPM} with a model that assumes two hotspots (\stu) in \Cref{sec:twohotspots}, which improves the residuals. Here we use two background approaches: \disc{} (\Cref{sec:stud}) and background marginalisation (\Cref{sec:stum}). We also test background marginalisation in the same section, and we test an alternative EoS-informed prior for mass and radius in \Cref{sec:EoS}. 

Supplementary materials are provided separately, containing complete corner plots (which display the \acp{PPD} for all parameters) for all model and data combinations to which we apply \ac{PPM} in this work. Also shown there are projection plots of the \acp{NS} with hotspot patterns for the \ac{MAP} samples.

\subsection{Single hotspot}\label{sec:singlehotspot}
\begin{figure*}
    \centering
    \includegraphics[width=0.9\textwidth]{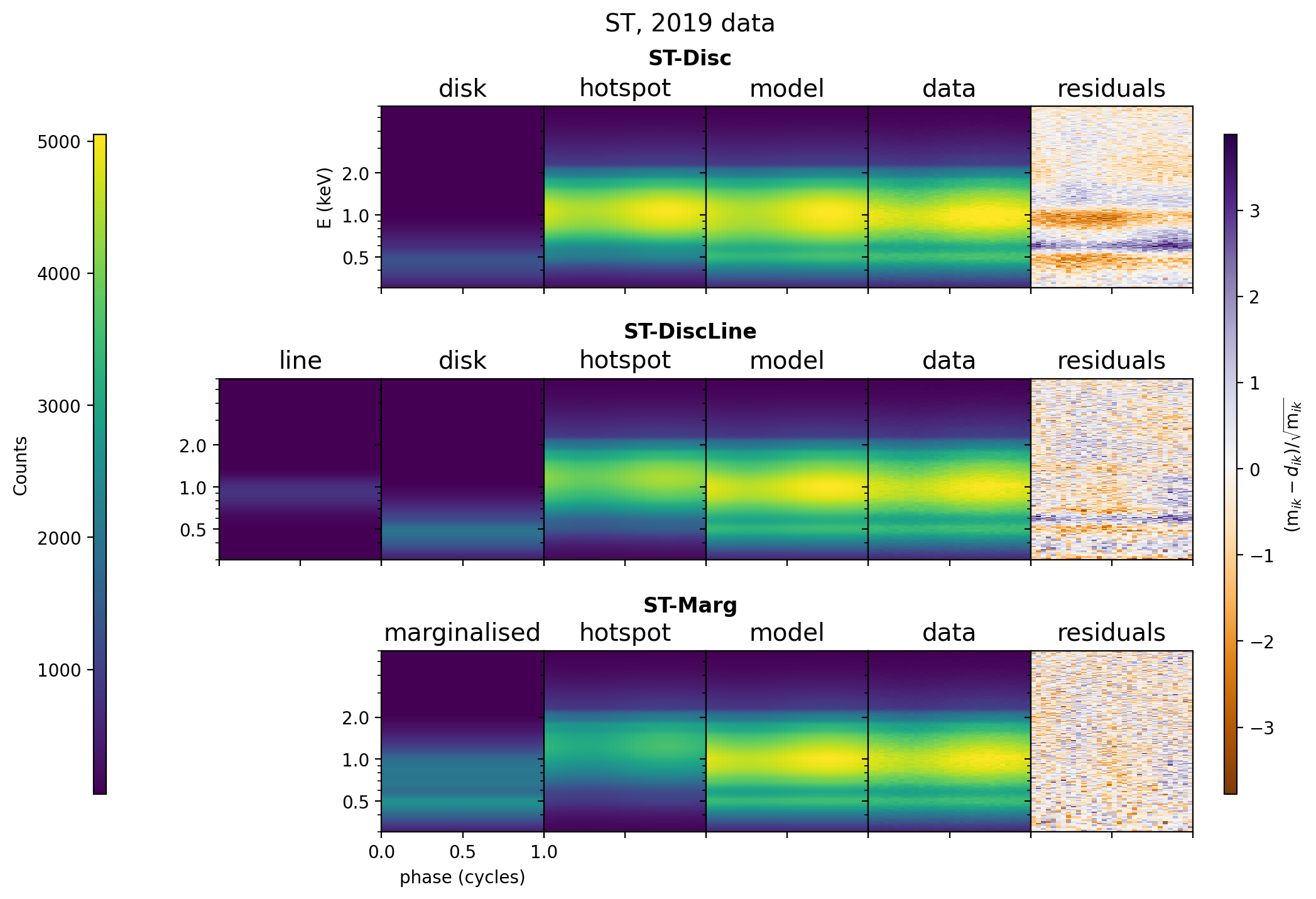}
    \caption{\ac{MAP} phase-energy resolved pulse profiles of various models. The panel titled `model' displays the expected pulse profile with the parameters of the \ac{MAP} sample. To the right the data and the right-most panel shows the normalized residuals between those two. In the panels left of the model, the decomposition into its constituent parts is shown. For all models, `hotspot' shows the pulse profile of the hotspot. In the left-most panel(s), the background contributions from various background models are shown: \disc\ (top), \discline\ (middle, separated) and the marginalised background (bottom).}
    \label{fig:ST_2019_plots}
\end{figure*}

This section describes the results of the \ac{PPM} with a single hotspot or \st\ model. Here, we combine the \ac{NS} model with three different approaches to background modelling: \std, \stdl, and \stm. \Cref{fig:ST_2019_plots} shows the modelled pulse profiles of the \ac{MAP} samples for the 2019 data, where each row shows the result for a different background approach. On the two right-most panels are the data and the residuals. The equivalent \Cref{fig:ST_2022_plots} for the 2022 outburst can be found in the Appendix. There are no notable differences in the results between the 2019 and 2022 data analysis with the single hotspot model. Here, we discuss the two datasets jointly. 

Focusing on the \std{} model first, \Cref{fig:ST_2019_plots,fig:ST_2022_plots} show that the \ac{MAP} sample of this model leaves a large residual with the data. In terms of spectral shape, the model is lacking a relatively narrow bump in counts at around 0.8--1.1~keV, as well as below around 0.5~keV, and overestimates counts between these bands. The hotspot produces the majority of the counts at 0.8--1.1 keV and is too spectrally broad to capture this feature in the data. There is also some phase dependence in the residual, which may indicate that there is some inaccuracy in the modelling of the pulsations. The full phase-energy resolved fit gives a $\chi^2/$DOF of 36709/18227 for the 2019 data and 33901/18227 for the 2022 data, both corresponding to a p-value of $<10^{-99}$. The p-value is the probability of observing a $\chi^2 > \chi^2_{\rm obs}$ given the model posterior expected counts (based on 100 draws from the posterior distribution). We conclude that this model does not explain the data.

Since there is a possibility that a spectral line feature (also found by \citealt{Sharma2023}) could be the cause of the feature at 0.8--1.1~keV, we added the Gaussian line (\Cref{eq:line}) to the model: \stdl. For the \ac{MAP} sample, we see that the spectral fit is improved, but it still leaves a very narrow negative spectral line at the interface between the disc and the line at around 0.6 keV, indicating that this background model consisting of a blackbody disc and broadened line is too simplistic to account for the data. \Cref{sec:discussion} discusses potential improvements to this model. There is some improvement in the phase-dependent residual, although some clustering is left, notably above 0.6 keV. As in with \std, this could point at some shortcoming in the modelling of the pulsations. The fit gives a $\chi^2/$DOF of 21920/18224 for the 2019 data (corresponding to a p-value \num{3.2e-74}) and 22462/18224 for the 2022 data (p-value of \num{2.4e-95}). We conclude that this model also cannot produce this data.

Thirdly, we employ the marginalised background model. This model has more flexibility and should therefore result in better spectral fits. And indeed, \Cref{fig:ST_2019_plots,fig:ST_2022_plots} show that this model is able to account for the spectral residual features. The $\chi^2/$DOF measure improves significantly at  19063/18227 (p-value of \num{8.2e-6}) for the 2019 data and 19010/18229 (p value of \num{2.7e-5}) for the 2022 data. Clustering in the energy-phase residuals is still visible. This gives us high confidence that even with an ideally flexible background, the single hotspot model cannot explain the data.

We also see that the marginalised background spectrum associated to the \ac{MAP} sample (meaning this background maximises the likelihood for the \ac{MAP} sample) in \stm{} is brighter than the \disc\ and the \discline. Consequently, the pulsations from the hotspot are dimmer. In the higher energy band, at above $\sim$2 keV, the marginalised background becomes restricted below the upper limit (discussed in \Cref{sec:model}) at $f_i*s$, where $s$ is the support factor, set to 100. Any potential bright phenomena above $\sim$2 keV are thus implicitly assumed to come from the hotspot. Because the background component nears the upper limit of its support boundary, we test the effect of varying $s$ on the \ac{PPM} analysis in \Cref{sec:stum}.

For all of these models, the residuals and fit statistics indicate that there are significant deficiencies in the modelling. Because of this, the inferred posterior distributions for the model parameters are not an authoritative indication for the system parameters, but they are nonetheless provided in \Cref{tab:ST_results} in the Appendix, with an overview of the fit statistics. The phase dependent residuals were present throughout all three background approaches, and because of this deficiency we next increase the complexity of the hotspot modelling by adding another hotspot.

\subsection{Two hotspots}\label{sec:twohotspots}
This section presents the results of the analysis with two hotspots in the \stu\ configuration: two single-temperature hotspots with unshared parameters. Similar to \Cref{sec:singlehotspot}, we discuss both 2019 and 2022 data sets jointly, but highlight noteworthy differences where present. 

We use two background approaches: the \disc\ model (\Cref{sec:stud}) and background marginalisation (\Cref{sec:stum}). The \discline\ model was omitted here due to computational limitations. This is the most expensive model to sample due to it having the highest number of parameters. In \Cref{sec:stum} we also investigate the robustness of the background marginalisation approach by varying the support boundary. In addition, we also analyse the data sets with a more restricted $M-R_{\rm eq}$ prior in \Cref{sec:EoS}, based on \ac{EoS} models and the results of \ac{PPM} analysis of other \acp{NS}. \Cref{tab:STU_results} shows an overview of the inferred parameters and fit statistics for all \stu\ model configurations. Evidence ratios are given compared to the \st\ variant of each model. Below we discuss the results of each model configuration separately.

\begin{table*}
\centering
\caption{\label{tab:STU_results} Overview of the posterior median values and 68 per cent \acp{CI} of all the parameters of all \stu\ model configurations. The header shows more information about each analysis: which \ac{NICER} dataset was analysed, the support factor $s$, live points, and the difference in log evidence compared to the equivalent \st\ run, $\chi^2$ fit metric of the posterior expected model counts, degrees of freedom and the p-value calculated from the two aforementioned metrics.}

\begin{tabular}{ccccccccc}
\hline
\hline
Model & \multicolumn{2}{c}{\stud} & \multicolumn{2}{c}{\stum} &  \multicolumn{2}{c}{\stum\ ($s$=1000)} & \multicolumn{2}{c}{\stume} \\
Dataset & 2019 & 2022 & 2019 & 2022 & 2019 & 2022 & 2019 & 2022 \\
$s$ & - & - & 100 & 100 & 1000 & 1000 & 100 & 100 \\
Live points & 2000 & 2000 & 2000 & 2000 & 1000 & 1000 & 1000 & 1000 \\
$\ln\Big(\frac{Z_{\rm STU}}{Z_{\rm ST}}\Big)^*$  & 2071 & 1139 & 303 & 321 & 1108 & 1130 & 248 & 317 \\
$\chi^2$  & 32551 & 31596 & 18409 & 18321 & 18335 & 18267 & 18528 & 18350 \\
DOF  & 18221 & 18221 & 18223 & 18223 & 18223 & 18223 & 18223 & 18223 \\
p-value  & $<$1e-99 & $<$1e-99 & 0.17 & 0.30 & 0.28 & 0.41 & 0.06 & 0.25 \\

&&&&&&\\

Parameters & & & & & & \\
\hline
$M\;\mathrm{[M}_{\odot}\mathrm{]}$ & $2.989_{-0.003}^{+0.002}$ & $2.779_{-0.005}^{+0.005}$ & $2.00_{-0.06}^{+0.06}$ & $1.93_{-0.09}^{+0.08}$ & $1.82_{-0.07}^{+0.06}$ & $1.79_{-0.07}^{+0.06}$ & $1.98_{-0.14}^{+0.12}$ & $1.46_{-0.17}^{+0.18}$\T\B \\
$R_{\mathrm{eq}}\;\mathrm{[km]}$ & $13.016_{-0.008}^{+0.009}$ & $12.09_{-0.02}^{+0.02}$ & $8.7_{-0.3}^{+0.3}$ & $8.3_{-0.4}^{+0.4}$ & $8.1_{-0.3}^{+0.3}$ & $7.8_{-0.3}^{+0.3}$ & $13.0_{-0.4}^{+0.3}$ & $12.4_{-0.3}^{+0.3}$\T\B \\
$D \;\mathrm{[kpc]}$ & $1.3207_{-0.0011}^{+0.0011}$ & $1.2870_{-0.0014}^{+0.0016}$ & $3.58_{-0.14}^{+0.11}$ & $3.15_{-0.17}^{+0.13}$ & $3.53_{-0.12}^{+0.11}$ & $3.24_{-0.13}^{+0.11}$ & $2.23_{-0.13}^{+0.15}$ & $2.70_{-0.19}^{+0.19}$\T\B \\
$N_\mathrm{H}\;[10^{21} \mathrm{cm}^{-2}]$ & $1.0767_{-0.0008}^{+0.0008}$ & $0.9565_{-0.0021}^{+0.0017}$ & $1.25_{-0.06}^{+0.06}$ & $0.91_{-0.07}^{+0.09}$ & $1.16_{-0.06}^{+0.06}$ & $0.82_{-0.07}^{+0.07}$ & $1.53_{-0.05}^{+0.06}$ & $1.12_{-0.07}^{+0.07}$\T\B \\
$i\;\mathrm{[deg]}$ & $29.60_{-0.02}^{+0.02}$ & $29.61_{-0.04}^{+0.04}$ & $70.7_{-3.0}^{+2.3}$ & $59.7_{-4.5}^{+3.3}$ & $73.2_{-3.6}^{+2.6}$ & $61.7_{-4.4}^{+4.4}$ & $30.7_{-0.8}^{+1.4}$ & $33.9_{-2.6}^{+3.5}$\T\B \\
$\phi_\mathrm{p}\;\mathrm{[cycles]}$ & $0.3153_{-0.0010}^{+0.0012}$ & $0.141_{-0.003}^{+0.004}$ & $0.101_{-0.007}^{+0.008}$ & $-0.018_{-0.015}^{+0.013}$ & $0.093_{-0.011}^{+0.013}$ & $-0.024_{-0.013}^{+0.013}$ & $0.050_{-0.011}^{+0.010}$ & $-0.087_{-0.017}^{+0.018}$\T\B \\
$\theta_\mathrm{p}\;\mathrm{[deg]}$ & $5.82_{-0.03}^{+0.03}$ & $5.20_{-0.10}^{+0.11}$ & $19.5_{-0.7}^{+0.7}$ & $23.9_{-1.3}^{+1.1}$ & $18.1_{-1.0}^{+0.8}$ & $23.9_{-1.2}^{+1.3}$ & $7.7_{-0.3}^{+0.3}$ & $6.4_{-0.5}^{+0.5}$\T\B \\
$\zeta_\mathrm{p}\;\mathrm{[deg]}$ & $47.13_{-0.05}^{+0.07}$ & $59.27_{-0.16}^{+0.18}$ & $84.2_{-3.3}^{+3.2}$ & $88.4_{-2.4}^{+1.1}$ & $79.9_{-2.9}^{+3.0}$ & $87.0_{-2.4}^{+1.8}$ & $31.9_{-2.4}^{+2.8}$ & $49.6_{-4.9}^{+4.8}$\T\B \\
$\tau_\mathrm{p}\;[-]$ & $2.630_{-0.003}^{+0.003}$ & $2.607_{-0.003}^{+0.003}$ & $1.37_{-0.02}^{+0.03}$ & $1.66_{-0.05}^{+0.05}$ & $1.317_{-0.016}^{+0.019}$ & $1.58_{-0.04}^{+0.05}$ & $1.54_{-0.02}^{+0.02}$ & $1.64_{-0.05}^{+0.06}$\T\B \\
$T_\mathrm{seed,p}\;\mathrm{[keV]}$ & $0.5369_{-0.0004}^{+0.0004}$ & $0.5445_{-0.0008}^{+0.0007}$ & $0.892_{-0.013}^{+0.013}$ & $0.914_{-0.015}^{+0.013}$ & $0.940_{-0.011}^{+0.010}$ & $0.965_{-0.015}^{+0.015}$ & $0.699_{-0.018}^{+0.016}$ & $0.640_{-0.018}^{+0.022}$\T\B \\
$T_\mathrm{e,p}\;\mathrm{[keV]}$ & $21.91_{-0.06}^{+0.07}$ & $20.56_{-0.05}^{+0.05}$ & $56.7_{-4.0}^{+5.0}$ & $39.3_{-2.9}^{+3.3}$ & $85.1_{-9.0}^{+9.0}$ & $47.1_{-4.7}^{+5.2}$ & $59.5_{-3.0}^{+3.6}$ & $41.1_{-2.8}^{+3.1}$\T\B \\
$\phi_\mathrm{s}\;\mathrm{[cycles]}$ & $0.5453_{-0.0004}^{+0.0004}$ & $0.400_{-0.002}^{+0.002}$ & $-0.054_{-0.005}^{+0.005}$ & $-0.189_{-0.006}^{+0.005}$ & $-0.059_{-0.004}^{+0.005}$ & $-0.194_{-0.005}^{+0.005}$ & $-0.083_{-0.005}^{+0.005}$ & $-0.231_{-0.005}^{+0.005}$\T\B \\
$\theta_\mathrm{s}\;\mathrm{[deg]}$ & $52.18_{-0.10}^{+0.12}$ & $124.1_{-0.3}^{+0.3}$ & $148.4_{-4.0}^{+2.9}$ & $139.3_{-5.5}^{+4.4}$ & $141.1_{-6.7}^{+4.1}$ & $132.0_{-6.6}^{+6.1}$ & $151.8_{-5.7}^{+4.6}$ & $146.3_{-4.7}^{+3.9}$\T\B \\
$\zeta_\mathrm{s}\;\mathrm{[deg]}$ & $5.835_{-0.015}^{+0.017}$ & $6.62_{-0.05}^{+0.06}$ & $48.8_{-3.5}^{+2.8}$ & $41.8_{-1.9}^{+2.2}$ & $47.5_{-4.3}^{+3.4}$ & $42.2_{-1.8}^{+2.5}$ & $38.0_{-7.4}^{+6.9}$ & $50.7_{-7.4}^{+6.1}$\T\B \\
$\tau_\mathrm{s}\;[-]$ & $0.507_{-0.003}^{+0.004}$ & $0.511_{-0.006}^{+0.010}$ & $0.69_{-0.11}^{+0.17}$ & $0.82_{-0.16}^{+0.18}$ & $1.33_{-0.11}^{+0.09}$ & $1.13_{-0.14}^{+0.12}$ & $0.89_{-0.13}^{+0.10}$ & $1.08_{-0.20}^{+0.18}$\T\B \\
$T_\mathrm{seed,s}\;\mathrm{[keV]}$ & $1.0248_{-0.0012}^{+0.0011}$ & $1.083_{-0.003}^{+0.003}$ & $0.683_{-0.020}^{+0.018}$ & $0.731_{-0.022}^{+0.018}$ & $0.74_{-0.02}^{+0.02}$ & $0.785_{-0.018}^{+0.019}$ & $0.517_{-0.004}^{+0.007}$ & $0.526_{-0.010}^{+0.017}$\T\B \\
$T_\mathrm{e,s}\;\mathrm{[keV]}$ & $28.4_{-1.2}^{+1.1}$ & $21.1_{-0.4}^{+0.5}$ & $58.9_{-18.6}^{+21.8}$ & $35.8_{-9.0}^{+11.6}$ & $22.6_{-1.6}^{+2.8}$ & $23.2_{-2.0}^{+4.0}$ & $24.4_{-2.6}^{+4.5}$ & $27.6_{-4.9}^{+8.7}$\T\B \\
$R_\mathrm{in}\;\mathrm{[km]}$ & $39.651_{-0.022}^{+0.015}$ & $38.64_{-0.05}^{+0.04}$ & - & - & - & - & - & -\T\B \\
$T_\mathrm{in}\;\mathrm{[keV]}$ & $0.10299_{-2e-05}^{+2e-05}$ & $0.11806_{-6e-05}^{+5e-05}$ & - & - & - & - & - & -\T\B \\
\hline
    \end{tabular}
    \vspace{2 mm}
    \begin{flushleft}
    \footnotesize{$^*$ For legibility, the increase in log evidence compared to the corresponding \st\ runs have been displayed, rather than the log evidence values individually. For \stud, the corresponding log evidences are from \std. For all \stum\ and \stume\ runs, the corresponding log evidences are from \stm. Because the \disc\ and marginalised background employ different log likelihood functions, the evidence values cannot be compared between \disc\ and marginalised background approaches. Evidences are also specific to their respective datasets and cannot be cross-compared.}\\
    \end{flushleft}
\end{table*}

\subsubsection{Disc background: \stud}\label{sec:stud}
\begin{figure*}
    \centering
    \includegraphics[width=0.9\linewidth]{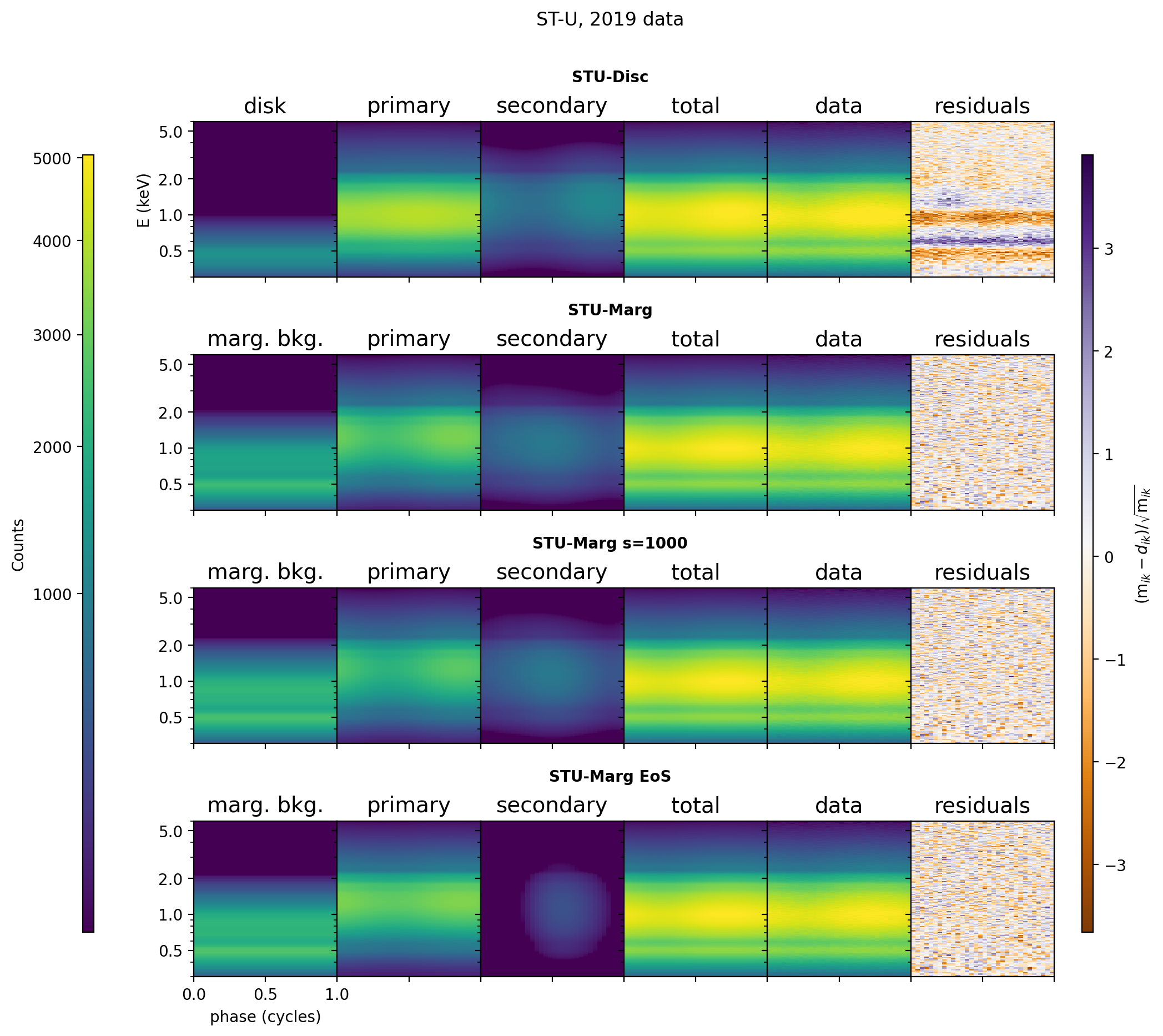}
    \caption{Data and modelled pulse profiles of the \ac{MAP} samples, where the pulse profiles are also decomposed in their contributions from primary and secondary hotspots, and background contributions.}
    \label{fig:STU_decomposition_2019}
\end{figure*}

\begin{figure*}
    \centering
    \includegraphics[width=0.9\textwidth]{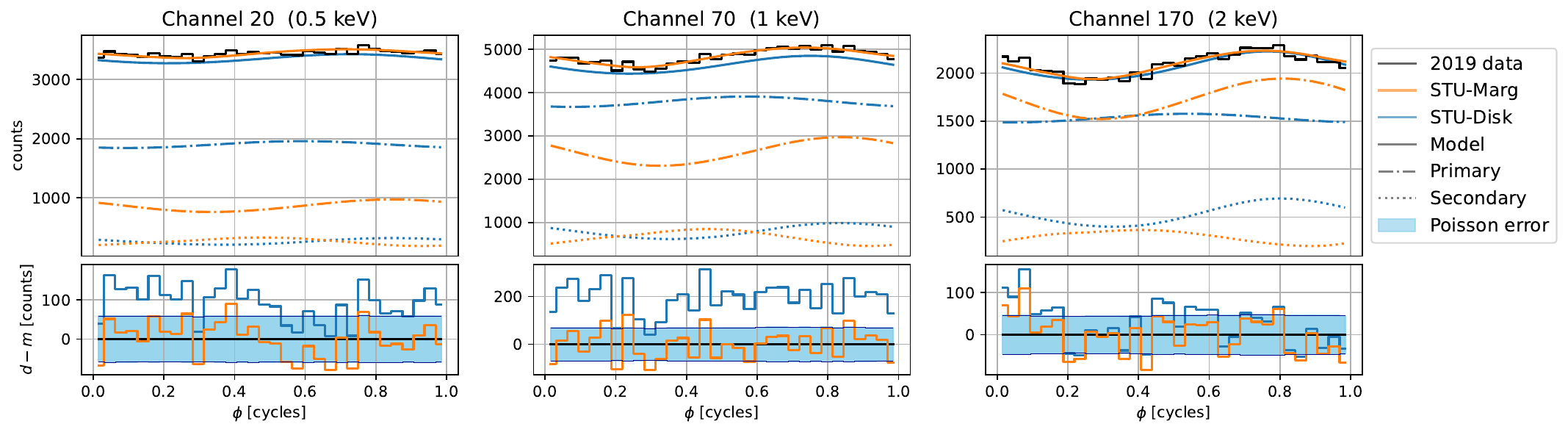}
    \caption{Pulse profiles of the \ac{MAP} samples in three representative \ac{NICER} channels for the \stud\ and \stum\ configurations. In the top panels, the data (in black) and total posterior predicted pulse profiles (i.e. spots and either disk or marginalised background contribution, solid lines) are shown along with the separate contributions of the primary (dash-dotted lines) and secondary hotspots (dotted lines). In the bottom panels, the residual counts (data-model) are shown in each representative channel, with the $\pm1\sigma$ Poisson errors.}
    \label{fig:STU_representative_2019}
\end{figure*}

As can be seen in \Cref{tab:STU_results}, including the second hotspot leads to a highly significant increase in evidence for both datasets. \Cref{fig:STU_decomposition_2019} shows the breakdown into model components, and \Cref{fig:STU_representative_2019} shows the same at three representative NICER channels at 0.5, 2 and 1 keV. The top-right panel of \Cref{fig:STU_decomposition_2019} and bottom panels of \Cref{fig:STU_representative_2019} show the residual between the \stud\ model and the 2019 data, which is still a bad fit despite the second hotspot. The $\chi^2/$DOF is 32551/18221 (p-value of $<10^{-99}$).

While residual features in the phase direction are at least consistent, now forming horizontal bands, the residual features in the spectral direction remain significant. From this it becomes clear that the two hotspots can account well for the pulsation, but are too spectrally broad to account for the spectral features in the data. Compared to \std, the primary hotspot contribution is more phase-independent, and the very small secondary hotspot introduces the right phase dependence to the overall pulse profile. The disc contributes to the radiation below 1~keV but only the hotspots at 1~keV and above. The equivalent plots for the 2022 data, \Cref{fig:STU_decomposition_2022,fig:STU_representative_2022}, are displayed in the appendix and the above statements hold true there as well. The only notable difference is that the fit to that data leads to the disc being brighter. In that case $\chi^2/$DOF is 31596/18221, corresponding also to a p-value of $<10^{-99}$. 

The accretion rates corresponding to the inferred $M, R_{\rm in}, R_{\rm eq}$, and $T_{\rm in}$ \citep[equation 3.23 in][]{Pringle1981} are $\sim1.9\times10^{-12}\,\rm M_{\odot}/yr$ and $\sim3.3\times10^{-12}\,\rm M_{\odot}/yr$ for the 2019 and 2022 data respectively. These rates are much lower than expected from \ac{J1808} based on independent estimates, for example $1.4-1.6\times10^{-10}\,\rm M_{\odot}/yr$ \citep{Casten2023}, based on the first hydrogen-triggered X-ray burst during the 2019 outburst.

Alongside the bad fits, we also see some inferred parameters approaching the edges of the priors, with many parameter values being similar to \std\ (\Cref{tab:ST_results}). Near the upper limit of the prior are $M$ and $R_{\rm in}$, and near the lower limit are $D$ and $i$. The \acp{CI} appear much smaller than we expect from \citetalias{Dorsman2025}. This is likely produced by a sharp decline in likelihood surface on one side of the posterior, and the prior edge on the other.

Overall, the bad fits and inferred parameters at edges of the priors signal that a major piece of physics is missing in this model. \Cref{sec:discussionbackground} discusses recommendations for accretion disc modelling.

\subsubsection{Marginalised background: \stum}\label{sec:stum}
The fit of \stum\ is much better, as shown in the lower right-most panels of \Cref{fig:STU_decomposition_2019,fig:STU_decomposition_2022} and bottom panels of \Cref{fig:STU_representative_2019,fig:STU_representative_2022}. 
The $\chi^2/$DOF for the 2019 data is 18409/18223 (p-value of 0.17) and for the 2022 data is 18321/18223 (p-value of 0.30), indicating that these models could reasonably produce this data. An improvement was already achieved with the \stm\ model over the \std\ model, but with \stum\ the phase dependence of the model is improved further. Compared to \stud, the background is much brighter, leading to dimmer pulsed radiation from the hotspots. This is the same effect that was seen when going from \std\ to \stm. Also notable is how the pulse profile of the individual hotspots is quite different, with the primary being dimmer and with a higher pulse fraction, while the secondary is offset in phase compared to \std. All the above statements are also true for the 2022 data, with the only notable difference being a brighter background.

\Cref{tab:STU_results} also shows that the inferred \ac{NS} parameters with this model have shifted significantly compared to the parameters also present in \stm. For both datasets, $M$ is now inferred to be around 2 $\mathrm{M}_\odot$ and the $R_{\rm eq}$ is small at 8--9 km. While for \std\ and \stud\ the star is viewed from near the pole, for \stm\ and \stum, it is viewed more from near the equator. For \stum, there are two similarly sized large hotspots in opposing hemispheres near the rotational axis. This contrasts against the single (very) large hotspot, covering around half the star, centred around the rotational axis. The overall result for \stum, with the reduction in size of the primary hotspot and shift in viewing angle, is that much of the constant component of the pulse profile of the primary has now been offloaded to the background. These results highlight the potential usefulness of tight priors on inclination, hotspot colatitude and angular radius.

\begin{figure}
    \centering
    \includegraphics[width=\linewidth]{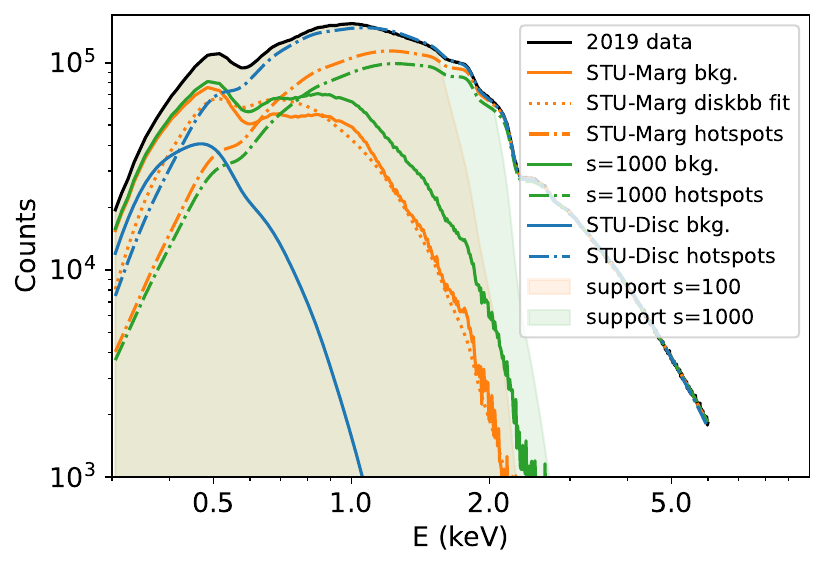}
    \caption{Phase-summed spectra of the data and inferred \stu\ components: backgrounds (solid lines) and combined hotspots (dash-dotted lines). In black is the spectrum from the 2019 \ac{NICER} data of \ac{J1808}. In orange inferred components of the \stum\ analysis with $s=100$. The dotted line is a \texttt{diskbb} fit to the inferred marginalised background. In green are the inferred components of the \stum\ analysis with $s=1000$. The shaded backgrounds indicate the region allowed by the support boundaries in the background marginalisation. The blue are the inferred components of the \stud\ analysis.}
    \label{fig:spectra2019}
\end{figure}

We study the spectral shape of the marginalised background by fitting \texttt{diskbb} to it. \Cref{fig:spectra2019} shows the phase-summed spectrum of the 2019 data in solid black, along with the marginalised background in solid orange. We find a best fit with the \texttt{diskbb} model to that background shown as the orange dotted line. The best fit is given by $T_{\rm in}=0.27$ keV and $R_{\rm in}=27$ km, which are reasonable parameters within the priors defined in \Cref{tab:parameters}. The corresponding accretion rate is still low, but closer to the value expected for \ac{J1808} at $\dot M = 4.3\times10^{-11}\,\rm M_\odot/yr$. While the best fit is good above 1.1~keV, the deviations below that energy are significant and resemble the residuals found from the \stud\ analysis. This deviation confirms that \texttt{diskbb} alone is not a sufficient model to fully capture the shape of the low-energy component, underlining the need for improved physics in the modelling. We discuss potential model improvements in \Cref{sec:discussion}. The analogous \Cref{fig:spectra2022} for the 2022 data is given in the Appendix, and the fit there leads to similar results: $T_{\rm in}=0.27$ keV, $R_{\rm in}=24$ km, and $\dot M = 3.3\times10^{-11}\,\rm M_\odot/yr$.

The methodology employed to infer the background in the \stum\ model takes as input a user-defined support factor $s$. Initially we had chosen an arbitrarily large value $s=100$. However, even with this large value the inferred background is restricted above around 2~keV. Therefore, it is likely that 100 was not large enough to be unrestrictive. 

\Cref{fig:spectra2019,fig:spectra2022} present how the inferred background is affected when $s$ is increased to 1000 for the 2019 and 2022 data respectively. For both data sets, even though the marginalised backgrounds approach the upper boundary only at relatively high energy ($\sim$2~keV) where there are relatively few counts, it is still clear the the increase of $s$ from 100 to 1000 has a significant effect on the marginalised backgrounds. 

\Cref{fig:STU_decomposition_2019,fig:STU_decomposition_2022} show that the overall fit quality improves slightly by eye. The $\chi^2/$DOF values improve: 18335/18223 (p-value of 0.28) for the 2019 data and 18267/18223 (p-value of 0.41) for the 2022 data. \Cref{tab:STU_results} also shows that the evidence is significantly better, and that the parameter values also make notable shifts (beyond 68 per cent CIs). For example $M$ decreases from $2.00^{+0.06}_{-0.06}$ to $1.82^{+0.06}_{-0.07}$ $\mathrm{M}_\odot$, and $R_{\rm eq}$ decreases from $8.7^{+0.3}_{-0.3}$ to $8.1^{+0.3}_{-0.3}$ km. This brings to light a dependency of the results on this value for $s$. \Cref{sec:discussion} discusses the implications of these results and recommendations for the approach for background modelling going forward.

\subsubsection{EoS informed prior and marginalisation background: \stume} \label{sec:EoS}
All the analyses mentioned so far explore the whole $M-R_{\rm eq}$ space with wide priors in $M$ and $R_{\rm eq}$. Given that the inferred radii with \stum\ were small, it is worth asking whether bad fits would be obtained if radii were restricted to the range inferred from contemporary \ac{EoS} theory and measurements. To address this question, this section shows the results of the \stume{} analysis, which is the same as \stum{}, but we include \ac{EoS} information in the prior for $M$ and $R_{\rm eq}$ by taking the \ac{EoS} inference results of \cite{Rutherford2024} as discussed in \Cref{sec:prior}. 

To start, we confirm that, as hypothesised, the sampling process required fewer computational resources. We find that for example the \ac{EoS}-informed run with 2019 data is of lower computational cost by a factor of $\sim 2.5$ (i.e., reduced to 8000 core-hours compared to 20000).

Next, to investigate the change in fit quality, the \ac{MAP} pulse profiles and residuals for this analysis are shown in the bottom panels of \Cref{fig:STU_decomposition_2019,fig:STU_decomposition_2022}. Focussing on the 2019 data first, we find that the $\chi/$DOF fit metric has worsened to 18528/18223 (p-value of 0.06), and that the \stume\ analysis yields moderately worse log-evidence values (54 in ln-space). Visually there are no clear changes in the residuals. This result slightly favours the usage of the \stum\ model, meaning it has a slightly better capability to fit the data well. For the 2022 data however, the worsening of the $\chi/$DOF metric is even smaller, to 18350/18223 (p-value of 0.25), and this time the log-evidence marginally decreases (4 in ln-space). Overall, these results do not robustly indicate that either model should be preferred, or that the usage of the \ac{EoS}-informed prior reduces the fit to the data.

Next, we report the inferred parameter values. We see in \Cref{tab:STU_results} that for the 2019 data, the inferred $R_{\rm eq}$ values are now higher, going from around $8-9$ km, to around $12-13$ km. While the inferred $M$ is roughly the same at $1.98_{-0.14}^{+0.12}$ M$_\odot$, it has shrunk significantly for the 2022 data to $1.46_{-0.17}^{+0.18}$ M$_\odot$ and is now in tension with the mass inferred from the 2019 data. Also, for both datasets the preferred geometries have shifted. The viewing angle is now closer to the rotational axis, similar to the \stud\ result. For both datasets the hotspots are still on opposing hemispheres. For the 2019 data, both hotspots have shrunk significantly ($\zeta_{\rm p}$ down by $\sim52^\circ$, $\zeta_{\rm s}$ down by $\sim11^\circ$), while for the 2022 data the primary shrinks (by $\sim39^\circ$) and the secondary grows (by $\sim9^\circ$). Based on both inferred geometries and broken-down pulse profiles, it appears that a separate solution is found compared to the \stum\ solution. Here, the secondary hotspot becomes hidden for a portion of the rotation. This is compensated by the primary hotspot which is always in view now and produces a lower pulse amplitude.

\begin{figure*}
    \centering
    \includegraphics[width=0.9\textwidth]{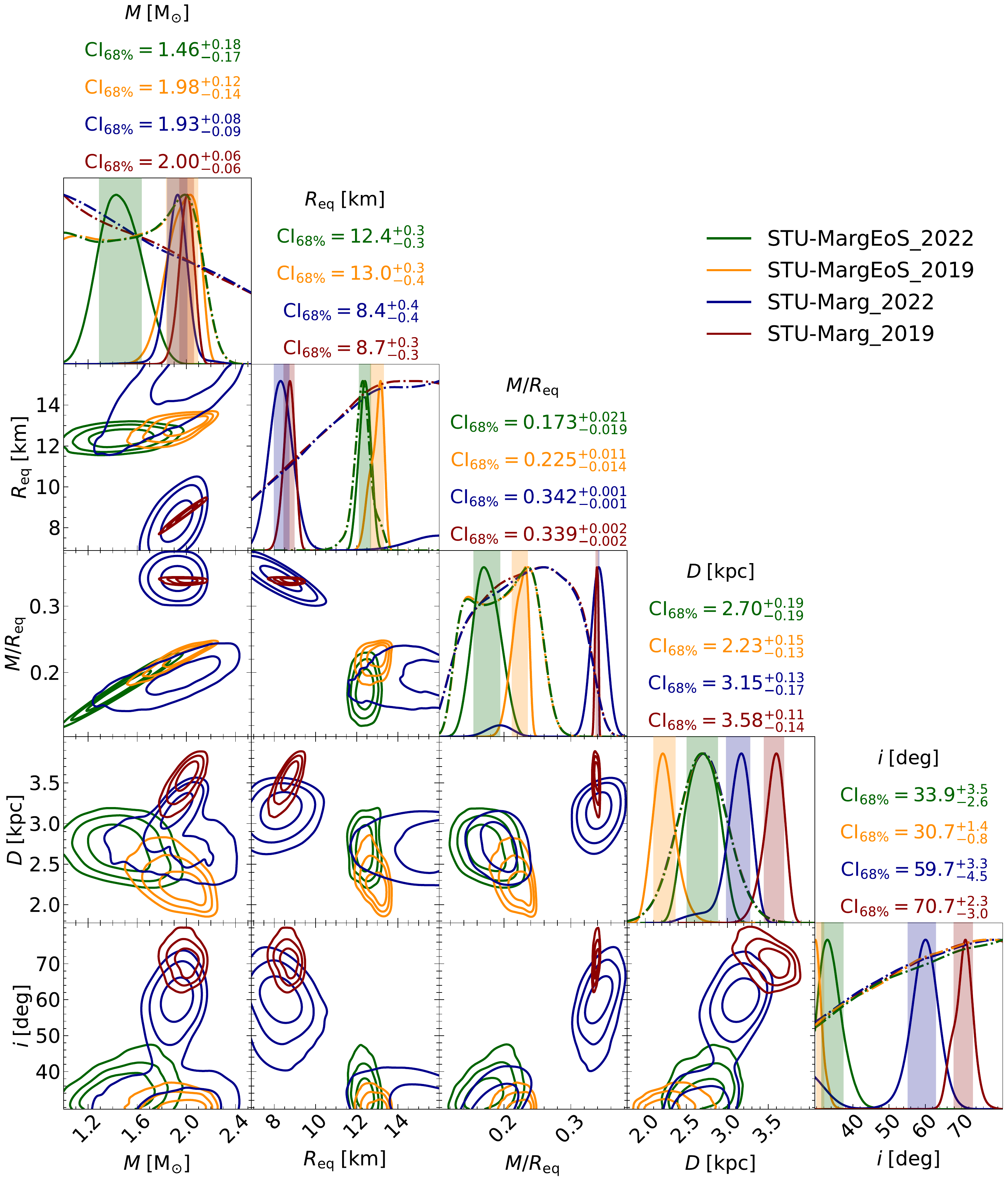}
    \caption{Marginal posterior distributions of the \stum\ ($s=100$) and \stume\ analyses for both 2019 and 2022 datasets of parameters that should be  consistent between the 2019 and 2022 outbursts of \ac{J1808}. Listed parameter values above the diagonal are median values along with 68 per cent \acp{CI}. On the diagonal are marginal posterior distributions for each parameter (solid lines), and prior distributions (dash-dotted lines). In the bottom triangle are 2D marginal posterior distributions of all pairwise combinations of parameters. Elongated contours are indicative of degeneracies between parameters. Contours in the 2D plots are the 68, 95 and 99.7 per cent credible levels.}
    \label{fig:STU_cornerplots}
\end{figure*}

A subset of inferred parameters should be consistent between outbursts and models. \Cref{fig:STU_cornerplots} shows the posterior distributions from the \stum\ and \stume\ analyses (with $s$=100), of these parameters: mass $M$, equatorial radius $R_{\rm eq}$, compactness,\footnote{Displayed as $M/R_{\rm eq}$, but it is the gravitational radius for mass $M$, $GM/c^2$, divided by $R_{\rm eq}$.} distance $D$, and inclination $i$. The \stud\ analysis has been omitted from discussion here because the residuals are too large and model improvement is required.

Rather than being mutually consistent, the posterior distributions appear to be separated into two modes, mode 1 has small $R_{\rm eq}$ (high compactness) and mode 2 has large $R_{\rm eq}$ (low compactness). The separation in these modes is also visible in the inclination and a subset of primary hotspot parameters: mode 1 has high inclination, large $\theta_\mathrm{p}$, large $\zeta_\mathrm {p}$ and large $T_{\rm seed}$. For the remaining parameters this separation is not (clearly) visible. Especially for the secondary hotspot the posteriors are wide and overlapping. While the \stum{} run with 2019 data only has significant posterior probably in mode 1, the 2022 data shows significant posterior probability for both modes. Both \stume{} runs select mode 2 due to their radius priors being confined to larger radius. 

The inference of $M$ is not divided by the two modes and we see a broad allowed distribution between ~1.3 and ~2.1 $M_\odot$. $D$ and $N_{\rm H}$ occupy a broad range covering much of their Gaussian priors. For both of them there is a weak divide between mode 1 and 2, meaning that a more precise estimate of both of these parameters would be somewhat helpful in distinguishing between the two modes. The parameters $i$, $\theta_{\rm p}$, and $\zeta_{\rm p}$ occupy a broad range as well, but for those parameters there is a strong divide between mode 1 and 2, meaning constraints on those parameters would be very effective in distinguishing between the two modes. $\phi_{\rm p}$, $\tau_{\rm p}$, $T_{\rm e, p}$ are tightly confined compared to their priors, and are not split along the two modes. $\tau_{\rm p}$, $T_{\rm e, p}$ show some slight mutual degeneracy. Of the parameters of the secondary hotspot, $\theta_{\rm s}$, $\zeta_{\rm s}$ $T_{\rm e,s}$, and $\tau_{\rm s}$ are overlapping, while $\phi_s$ and $T_{\rm seed,s}$ are split between modes 1 and 2.

\section{Discussion}\label{sec:discussion}
\subsection{Key findings and implications}
We analysed the NICER data of the 2019 and 2022 outbursts of \ac{J1808}, carrying out \ac{PPM} for the accretion-powered pulsations in the peak of the outburst. We find that a single hotspot model is insufficient to fit the data for all three approaches considered here for modelling the accretion disc. A two-hotspot model still did not explain the data well when fitting the accretion disc with a physical model, \texttt{diskbb}. This led to spectral residual features below around 1.1\,keV. The peak at 0.8--1.1\,keV could correspond to the presence of (a blend) of broadened reflection lines off the inner disc edge. That result implies more physics is needed in the accretion disc model to properly account for the data. 

The two-hotspot model is able to explain the data well if a very flexible marginalised background approach is used to account for the accretion disc background. However, even then the inferred background spectrum and \ac{NS} parameters, including mass and radius, are found to cover a wide range depending on the user-specified arbitrary upper limit for background counts. No physics has yet been used to inform these upper limits, and more work is needed to define appropriate limits. 

An exploratory run was also done with the same background approach but with a constrained prior on mass and radius based on dense matter theory and measurements. This shifted inferred parameters, but did not significantly reduce fit quality or model evidence, suggesting that the \ac{J1808} data is not in tension with previous \ac{EoS} constraints.

Taking all these points together, the reported inferred parameters here are not final but are subject to further refinement pending model improvements. In the sections below, \Cref{sec:discussioncomparison} compares the results obtained here to previous results by \citetalias{Salmi2018}. \Cref{sec:discussionbackground,sec:discussionhotspots,sec:discussioncaveatsrecommendations} discuss further caveats and recommendations, including those already touched upon above, in more detail.

\subsection{Comparison to Salmi et al. 2018}\label{sec:discussioncomparison}
\citetalias{Salmi2018} use a similar Bayesian framework to analyse pulse profiles to put parameter constraints on \ac{J1808}. They used \ac{RXTE} data from 1998 with energy channels ranging from 3 to 18 keV, divided into 16 phase bins and 24 energy channels. At this higher energy, the disc blackbody contribution is small, so they omit this component. This alleviates the degeneracy between disc blackbody and \ac{NS} that we had to contend with in modelling \ac{NICER} data.

However, unlike in this study, they did need to account for higher energy phenomena. These are an iron line at 6.4 keV as well as the Compton reflection continuum above 10 keV. They obtained moderate fits with the \ac{J1808} data and retrieved a somewhat low value for $M = 1.13^{+0.13}_{-0.06} $M$_{\odot}$ and a very low value $R_{\rm eq}=5.8^{+0.9}_{-0.7}$km, which they noted is outside expectations for modern \acp{EoS}. They also present a grid of posterior distributions for various parameters conditional on a grid with fixed mass values. Their results are somewhat compatible with the high radius mode found here, but highly in tension with the low radius mode. In addition, their inferred inclination stays at around 80 degrees regardless of the mass. Also, above 1.8 M$_\odot$, their inferred distance shoots quickly up to their upper boundary at 6 kpc. Our inferred parameters are in tension with these results for both modes. Our low radius mode fits better with their inferred distance and inclination, but its high mass is in strong tension. Our high radius mode is compatible with their mass and radius, but prefers a lower inclination. Taken together, these discrepancies suggest more work is needed on the modelling and that high energy data would place valuable complementary constraints on mass and radius estimation with the \ac{PPM} technique for \acp{AMP}. These results also underscore the importance of robust inclination estimates for mass and radius estimation.

It is also notable that \citetalias{Salmi2018} find much smaller hotspots across their mass grid, with angular radii ranging between 10 and 30 degrees. Although it should be noted that their prior upper limit was smaller at 40 degrees, this still suggests that simultaneous high energy data alongside low energy data could play a role in pinning down the relative contribution of the disc versus \ac{NS} hotspots in the non-pulsed low energy component.

\subsection{Accretion disc (background) modelling}\label{sec:discussionbackground}
Modelling the background with a blackbody disc model left similar looking spectral residuals with both one and two hotspots. These are a relatively narrow bump in the data at around 0.8--1.1~keV, as well as around 0.5~keV. It turned out this spectral shape is too narrow to be reproduced by a combination of a disc blackbody and \ac{NS} hotspot(s). 

A deviation from the assumed solar elemental abundances in the line-of-sight absorption, represented in the effective $N_{\rm H}$, could cause this deviation. However, spectral fitting of the phase-averaged spectrum with \textsc{XSPEC}, freeing the elemental abundances of O, Fe and Ne (which have their absorption edges in this energy range), led to only marginal improvement and did not resolve the residuals. On top of this, \cite{pinto2014} studied the absorption lines of \ac{J1808}, and found the elemental abundances in the line-of-sight to be consistent with solar abundances.

The residuals could be caused by broadened reflection lines \citep{Sharma2023, Chakraborty2024}, so we also modelled the background as a disc blackbody with a Gaussian line feature. We only tried this with one hotspot due to limited computational resources.\footnote{The limitation is caused by the increase in the number of parameters, which increases the sampling cost.} This improved the residuals moderately, but still left a minor spectral feature at the interface of the disc and line components at 0.6~keV. 

This motivated the use of the final background approach: background marginalisation. This approach is very flexible: there is no underlying physics model required and the background counts in each energy channel are independent from each other. As expected, we found a significant improvement in the spectral fit, but with one hotspot some phase-dependent residuals still remained. With two hotspots, the residuals improved yet further, with the data now being well accounted for in both the spectral and phase dimensions.

However, it is important to note that the inferred marginalised background depends on a pre-defined lower and upper boundary. The boundaries were defined as [$f/s, f*s$], where $s$ is a multiplicative support factor, where $f$ is the flux of a fiducial disc model. Upon usage of a larger $s=1000$ instead of 100, we found a significant increase in the log evidence. However, it is not sure yet if this value would be sufficient (or too large) to represent the physical accretion disc flux.

Although the background only reached the upper boundary at $\sim2$~keV, the whole background spectrum was shifted depending on $s$, as shown by \Cref{fig:spectra2019,fig:spectra2022}. This shift is possible due to the fact that from \ac{NICER} pulse profiles alone, the accretion disc cannot be distinguished from the non-pulsed hotspot radiation. Hotspots can also produce a non-pulsed component, as long as at least some part of them is continuously in view during one rotation. This was the case in the obtained solutions here too, as shown in \Cref{fig:STU_decomposition_2019,fig:STU_decomposition_2022}. More accurate hotspot modelling, discussed in \Cref{sec:discussionhotspots}, could reduce the degeneracy here. In the end, both the choice in boundary $s$ and fiducial disc flux $f$ will rule out some parameter space, shifting the posterior distributions of parameters. Synthetic data studies could be done to quantify the shift of parameter posteriors and produce recommendations for fiducial disc models and background boundary.

Given the dependency of the results on $s$, it would be preferable not to use an arbitrary boundary. However, in preliminary tests with synthetic data, we found in cases where no support boundary is used and where $s$ is large (i.e. $s\gtrsim 100$), the analysis is prone to biases. Specifically, if in some channels the true background has a near-zero count rate, but a higher background is allowed, this typically leads to an overestimated background and biases to many of the star parameters. However, more testing is warranted, for example to analyse whether the usage of more live points could also resolve this bias. This has not been done yet due to the computational cost.

The best solution would be to use a fully physics-based accretion disc model. However, the model used so far produces large residuals, so some physics must be missing. The Gaussian line we used for one hotspot still left a small spectral feature, so even if a line feature is the right idea, this model is probably too simplistic. To account for this, using a more sophisticated disc reflection model such as \texttt{xilconv} as used by \citetalias{Salmi2018}, is probably warranted. We are also missing special and general relativistic effects in the disc \citep[e.g.][]{Loktev2022}. An alternative could be to cut the energy contribution below 0.6~keV, the point below which the problems arise. It is also possible that a second blackbody contribution -- plausibly from the non-hotspot surface of the NS -- could improve the fit with the data. Increasing complexity must be done with caution, however, because a more expensive model or an increase in the number of parameters increases the computational cost of nested sampling.

If this flexible background approach with boundaries cannot be avoided, a physics-informed boundary would at least be an improvement. One could define an upper limit with an accretion disc model using an estimation of the inner disc radius through the iron line from higher energy data with high spectral resolution, or an upper limit on the disc temperature through an upper limit on the accretion rate. Tighter bounds on the background are very helpful to reduce computational expense due to reduced exploration of parameter space required.

\subsection{Hotspot modelling}\label{sec:discussionhotspots}
Although modelling \ac{J1808} with a single circular hotspot is well-established \citep[e.g.][]{Poutanen2003, Salmi2018, Bobrikova2023}, it led to significant phase-residuals. The step up to two circular hotspots significantly improved the residuals, for all choices of background models. While it is thus clear that the surface pattern of \ac{J1808} is better described by two circular hotspots, one must be careful in the interpretation of these results. There is one case, the \stud\ model and the 2019 data, where the \ac{MAP} hotspot pattern consists of a large and small hotspot that are nearly touching. This resembles more a large single hotspot with a complex shape than two hotspots at opposing sides of the star. In all remaining cases the \ac{MAP} surface pattern does feature hotspots on (near) opposing sides of the star (see also the supplementary materials).

Focusing now only on the two-hotspot models, another finding is that the hotspots are consistently large. For \stud\ we find combinations of large and small hotspots, where the angular radius of the large hotspot is up to 60 degrees. For \stum\ both hotspots are on near opposing sides of the star and more similar in size, with the largest primary hotspot encompassing almost half of the star. For \stume\ the hotspots are again on opposing sides of the star and the angular radius of the largest hotspot has shrunk to around 50 degrees. While large hotspots are not in tension with results of (general relativistic) magneto-hydrodynamic modelling, the circular shapes are in tension with their more elongated and crescent-like shapes \citep{Romanova2004,Kulkarni2013,Das2025}. This is a motivation to transition to more complex shapes for future analyses.

As visible in \Cref{fig:STU_decomposition_2019,fig:STU_decomposition_2022}, the combination of the viewing angles and large circular hotspots enables a strong non-pulsed component from the star. These large shapes could thus be necessitated by the need for a non-pulsed high energy component if the background model does not provide it. Unfortunately, as also mentioned in \Cref{sec:discussionbackground}, the trade-off in non-pulsed radiation between background and hotspot is difficult to pin down with the \ac{NICER} data. Besides improved accretion disc background modelling, we also expect that improved modelling of \ac{NS} radiation, such as hotspot shapes and constraints on these shapes (and possibly radiative transfer in the accretion column,  \citealt{Ahlberg2024}), will provide more realistic fits of the \ac{NS} and accretion disc system. We also expect that polarization data could improve fits as it helps to estimate inclination, hotspot colatitude and hotspot size (if the hotspot spans a significant part of the \ac{NS} surface the polarization degree and angle will be affected). 

Another model addition to increase accuracy (and rule out incorrect parameter space) would be implementing light ray occultation caused by the disc. The systems that will be most constrained by disc occultation would be those with large inclinations. The contributions from the secondary hotspots would then be most affected (assuming they are on the opposing hemisphere with respect to the observer). However, by constraining the pulse profile of the secondary the overall signal is affected, and therefore the primary hotspot will also be constrained. 

\subsection{Further caveats and recommendations for future research}\label{sec:discussioncaveatsrecommendations}
This section lists further caveats and recommendations beyond the accretion disc (background) and hotspot modelling. To start, we refer the reader to section 6.4 of \citetalias{Dorsman2025}, because many of their caveats and suggested model improvements are still valid here. 

A first recommendation is the possibility for a joint Bayesian analysis of the 2019 and 2022 datasets, which has not been tried here due to limitations in computational resources. In that case, mass, radius, distance, inclination and $N_\mathrm{H}$ would be shared, while other parameters would vary between datasets. Similarly, joint analysis that includes observations made by other instruments, especially if simultaneous, could provide complementary constraints. For example, \texttt{AstroSat} also observed the 2019 outburst \citep{Sharma2023} and 2022 outburst \citep{Kaushik2025}. Higher energy data could provide complementary constraints on the hotspot geometry and atmosphere parameters.

A notable finding in this study was that the spectrum deviates from the assumed spectrum in \citetalias{Dorsman2025}. While it is possible that the 1~keV feature may be enhanced by an unexpected instrumental background or uncertainty in the energy dependent effective area, this is unlikely because it was also observed by \texttt{AstroSat} \citep{Sharma2023}. Given this example of a spectral deviation, we recommend an additional synthetic data study to quantify the effect on parameter inference with \ac{NICER} data due to unaccounted-for deviations, such as a Gaussian line feature, with all accretion disc modelling approaches. 

This study also tested the usage of \ac{EoS}-informed priors. This restriction in $M-R$ prior shifted \acp{PPD} to be more in line with modern \ac{EoS} results, but did not lead to any significant change in Bayesian evidence (at least for the 2022 data), indicating neither model is preferred. The required computational resources reduced significantly, by a factor of $\sim2.5$. Given these considerations, we consider that further testing with \ac{EoS} informed priors is worthwhile - particularly if we wish to explore more complex, and therefore more computationally expensive, surface patterns. 

We further note that this analysis was done on pulse profiles constructed by averaging over fairly long observations of 7 and 11 days, during which there both the flux varied significantly and a minor change in the fractional pulse amplitude of up to 4 per cent was visible. Breaking down the pulse profiles into smaller sections would enable tracking of the evolution of the accretion and hotspots throughout the outburst. However, we also note that this would represent an increase in necessary computationally resources, because more datasets would require more likelihood evaluations.

Finally, we note that \acp{AMP} have a rich phenomenology, and there are independent estimates of model parameters (notably mass, radius, distance and inclination, and $N_\mathrm{H}$, which must be consistent between observations) available that have not been incorporated in this study, such as the study of thermonuclear bursts (e.g. \citealt{Goodwin2019, Casten2023}, who studied this for \ac{J1808} specifically) and burst oscillations (e.g. \citealt{Kini2024b}, who studied XTE J1814-338). Accretion disc parameters can also be constrained through independent measurements such as mapping of the reverberation lag of kHz quasi-periodic oscillations \citep[see e.g.][who studied 4U 1728-34]{Coughenour2020}, burst-disc interaction \citep[see e.g.][for a review]{Degenaar2018}, and broadened Fe lines \citep[e.g.][who studied \ac{J1808}]{Papitto2009}. In future studies it would be interesting to cross-check constraints from PPM with constraints from other methods, or even to consider joint fitting.

\section{Conclusion}\label{sec:conclusion}
This study performed a Bayesian analysis of the \ac{NICER} persistent pulse profiles during the peaks of the 2019 and 2022 outbursts of SAX J1808.4$-$3658 with the aim of estimating model parameters. In initial fitting we tried a model with a single circular hotspot, as might be expected if the accretion disc obscures the view of the other hemisphere of the star. This model left significant energy and phase residuals, and is therefore insufficient to account for the data. A notable residual feature that resembled a broadened reflection line at 0.8--1.1 keV was found, and while accounting for it with a Gaussian line improved the fit, it still left both phase and energy residuals. Using background marginalisation, where the accretion disc model was replaced by a flexible background spectrum, still left noticeable phase residuals.

With a model containing two hotspots, we found that modelling the accretion disc with a disc blackbody model did not fit the spectrum well, leaving the line-like residual feature at 0.8--1.1 keV. This result indicates that even with two hotspots, the simple accretion disc model applied here does not fully account for the physics, motivating our recommendation to use more accurate accretion disc and reflection line modelling in future work. 

Use of background marginalisation led to significantly improved residuals. However, some portion of the unpulsed counts could be produced by the NS hotspots. Because of this degeneracy the background marginalisation shifts depending on the predefined background upper and lower boundaries, rendering the inferred values of model parameters, including mass and radius, less robust compared to physics informed background modelling. Polarimetric data, as well as higher energy data, are expected to play a complementary role in similar analyses by providing independent constraints on these model parameters, including inclination, hotspot colatitude, hotspot size as well as the inner accretion disc radius.

\section*{Acknowledgements}
B.D. thanks Nathan Rutherford for discussions of the EoS informed approach, Niek Bollemeier for assistance with \textsc{XSPEC}, and Duncan Galloway for discussions related to X-ray bursts and distance estimates. B.D., T.S., and A.L.W. acknowledge support from ERC Consolidator grant No. 865768 AEONS (PI: Watts). M.N. is a Fonds de Recherche du Quebec– Nature et Technologies (FRQNT) postdoctoral fellow.

This work was supported in part by NASA through
the NICER mission. This work was sponsored by NWO Domain Science for the use of supercomputer facilities. This work used the Dutch national e-infrastructure with the support of the SURF Cooperative using grant no. EINF-5867 and is subsidized by NWO Domain Science. Part of the work was carried out on the HELIOS cluster including dedicated nodes funded via the above mentioned ERC CoG. We acknowledge extensive use of NASA’s Astrophysics Data System (ADS) Bibliographic Services and the ArXiv.

\section*{Software}
\textsc{X-PSI} \citep[version 3.0.0, ][]{Riley2023}, GNU Scientific Library (\textsc{GSL}; \citealt{Gough2009}), \textsc{HEASoft} \citep{HEASoft}, \textsc{MPI} for Python \citep{Dalcin2008}, \textsc{Multinest} \citep{Feroz2009}, \textsc{Pymultinest} \citep{PyMultiNest}, \textsc{nestcheck} \citep{Higson2018JOSS}, \textsc{GetDist} \citep{Lewis2019}, \textsc{Jupyter} \citep{2007CSE.....9c..21P, kluyver2016jupyter}, \textsc{astropy} \citep{astropy:2013, astropy:2018, astropy:2022}, \textsc{scipy} \citep{2020SciPy-NMeth, scipy_10909890}, \textsc{matplotlib} \citep{Hunter:2007}, \textsc{numpy} \citep{numpy}, \textsc{python} \citep{python}, \textsc{Cython} \citep{cython:2011} and \textsc{spyder} \citep{Spyder}.

\section*{Data availability}
A basic reproduction package for the analysis, including all the data, analysis scripts, and all the figures is available at \href{https://doi.org/10.5281/zenodo.17232362} {10.5281/zenodo.17232362}.



\bibliographystyle{mnras}
\bibliography{bibliography} 






\appendix
\section{Inferred parameters for ST and figures for the 2022 data}
\begin{figure*}
    \centering
    \includegraphics[width=0.9\textwidth]{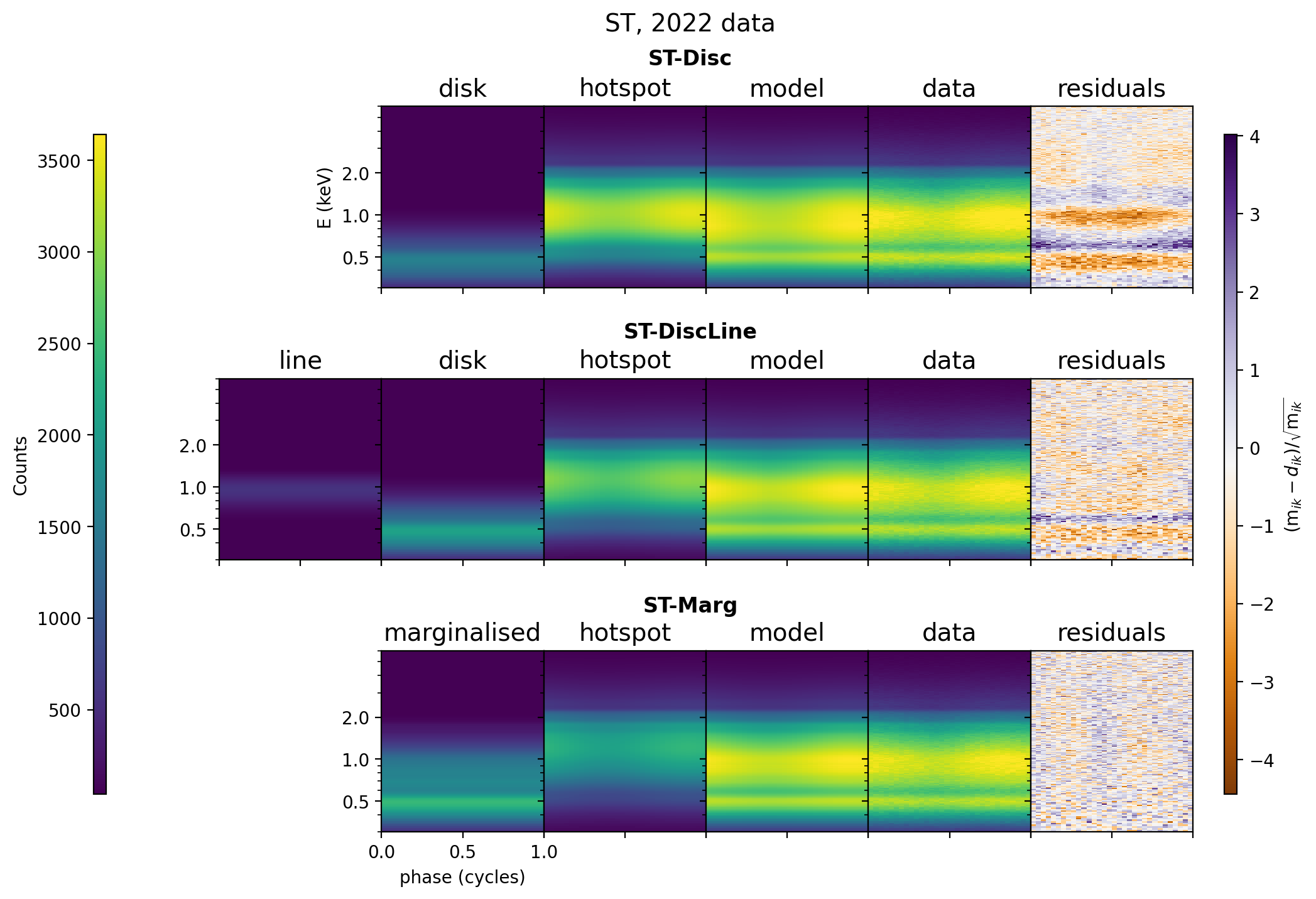}
    \caption{Decomposition of the \ac{MAP} pulse profiles as well as residuals for the \st\ models with 2022 data. See the caption of \Cref{fig:ST_2019_plots} for more details.}
    \label{fig:ST_2022_plots}
\end{figure*}

\begin{figure*}
    \centering
    \includegraphics[width=0.9\linewidth]{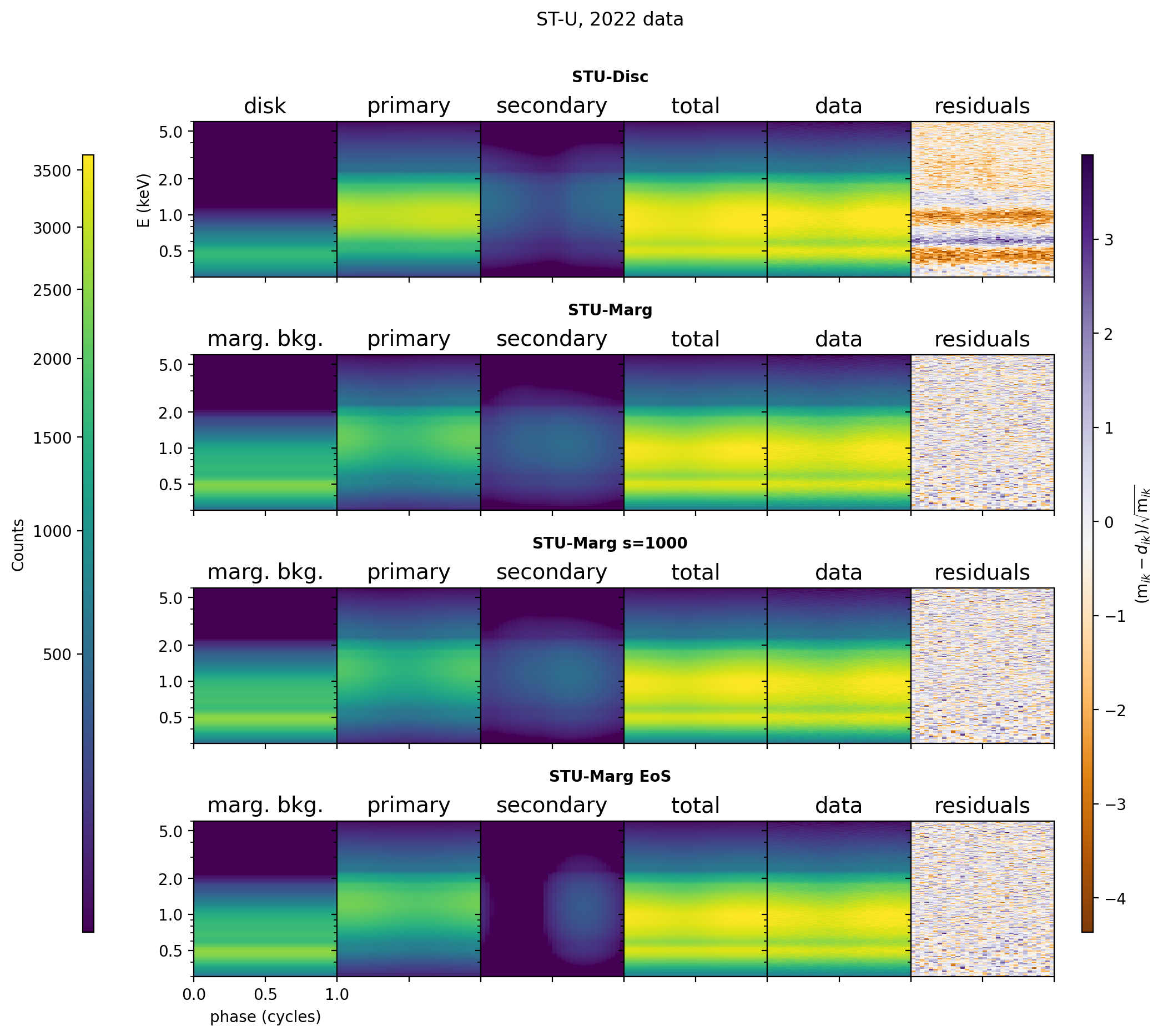}
    \caption{Decomposition of the \ac{MAP} pulse profiles as well as residuals for the \stu\ models with 2022 data. See the caption of \Cref{fig:ST_2019_plots} for more details.}
    \label{fig:STU_decomposition_2022}
\end{figure*}

\begin{figure*}
    \centering
    \includegraphics[width=0.9\linewidth]{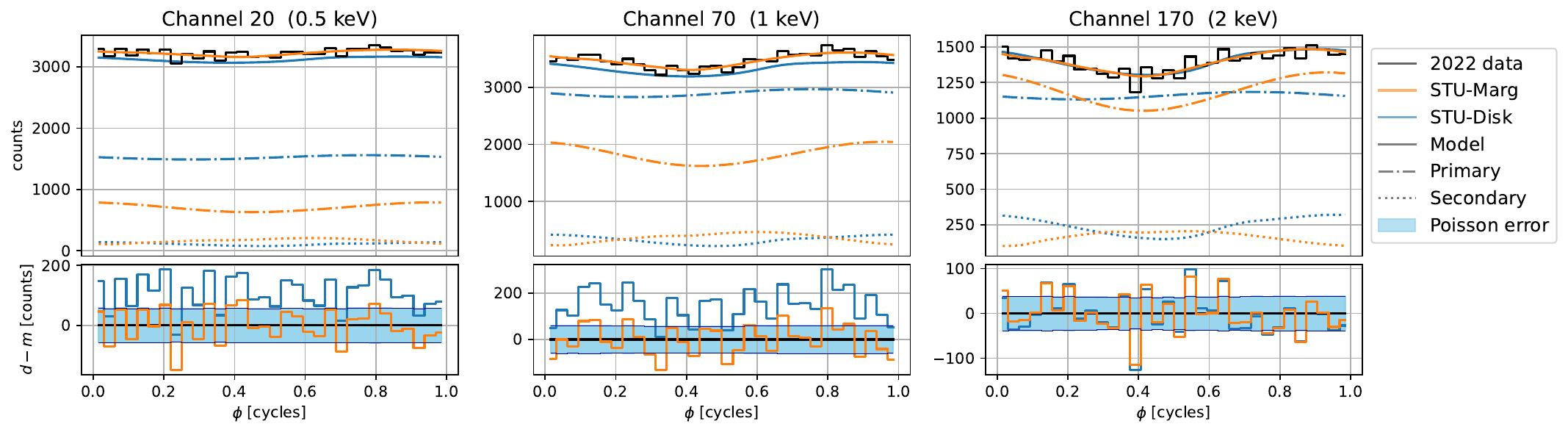}
    \caption{\ac{MAP} pulse profiles as well as residuals for the \stud\ and \stum\ configurations with 2022 data broken down into three representative \ac{NICER} channels. See the caption of \Cref{fig:STU_representative_2019} for more details.}
    \label{fig:STU_representative_2022}
\end{figure*}

\Cref{tab:ST_results} shows the median and 68 per cent \acp{CI} of the model parameters for the \st\ model in all configurations. \Cref{fig:ST_2022_plots,fig:STU_decomposition_2022,fig:STU_representative_2022} show the pulse profile decompositions and residuals for the \ac{NICER} data of the 2022 outburst of \ac{J1808}.

\begin{table*}
    \centering
    \caption{\label{tab:ST_results} Overview of the posterior distributions of all the model configurations with a single hotspot. For each model and corresponding parameters, we show the median and 68 per cent \acp{CI} of the marginalised posterior distributions. The header includes the $\ln{Z}$ evidence obtained for each model and the other elements are described in detail in \Cref{tab:STU_results}.}
\begin{tabular}{ccccccc}
\hline \hline 
Model & \multicolumn{2}{c}{\std} & \multicolumn{2}{c}{\stdl} & \multicolumn{2}{c}{\stm} \\
Dataset & 2019 & 2022 & 2019 & 2022 & 2019 & 2022\\
Live points & 4000 & 4000 & 4000 & 4000 & 4000 & 4000 \\
$\ln Z$ & 168817383 & 118452791 & 168824899 & 118458889 & $-$90996 & $-$87791 \\
$\chi^2$  & 36709 & 33901 & 21920 & 22462 & 19063 & 19010 \\
DOF  & 18227 & 18227 & 18224 & 18224 & 18229 & 18229 \\
p-value  & $<$1e-99 & $<$1e-99 & 3.2e-74 & 2.4e-95 & 8.2e-06 & 2.7e-05 \\

&&&&&&\\
Inferred parameters & & & & & & \\
\hline
$M\;\mathrm{[M}_{\odot}\mathrm{]}$ & $2.9985_{-0.0012}^{+0.0005}$ & $2.993_{-0.001}^{+0.001}$ & $2.998_{-0.002}^{+0.001}$ & $2.999_{-0.002}^{+0.001}$ & $1.02_{-0.01}^{+0.03}$ & $1.06_{-0.04}^{+0.09}$\T\B \\
$R_{\mathrm{eq}}\;\mathrm{[km]}$ & $15.118_{-0.018}^{+0.009}$ & $13.047_{-0.017}^{+0.009}$ & $13.06_{-0.02}^{+0.02}$ & $13.43_{-0.30}^{+0.05}$ & $7.6_{-1.4}^{+0.9}$ & $9.9_{-1.1}^{+1.4}$\T\B \\
$D \;\mathrm{[kpc]}$ & $1.5240_{-0.0004}^{+0.0015}$ & $1.360_{-0.002}^{+0.001}$ & $1.2013_{-0.0009}^{+0.0018}$ & $1.2004_{-0.0003}^{+0.0006}$ & $2.4_{-0.1}^{+0.2}$ & $2.5_{-0.2}^{+0.2}$\T\B \\
$N_\mathrm{H}\;[10^{21} \mathrm{cm}^{-2}]$ & $1.0434_{-0.0013}^{+0.0005}$ & $0.943_{-0.001}^{+0.001}$ & $1.258_{-0.002}^{+0.002}$ & $1.086_{-0.002}^{+0.002}$ & $1.32_{-0.06}^{+0.06}$ & $0.94_{-0.07}^{+0.07}$\T\B \\
$i\;\mathrm{[deg]}$ & $29.70_{-0.04}^{+0.07}$ & $29.67_{-0.05}^{+0.05}$ & $29.8_{-0.2}^{+0.3}$ & $32.8_{-0.2}^{+0.8}$ & $78.1_{-15.9}^{+2.4}$ & $78.9_{-3.3}^{+1.8}$\T\B \\
$\phi\;\mathrm{[cycles]}$ & $0.1703_{-0.0006}^{+0.0004}$ & $-0.0069_{-0.0005}^{+0.0006}$ & $0.135_{-0.001}^{+0.001}$ & $0.014_{-0.002}^{+0.002}$ & $0.226_{-0.001}^{+0.001}$ & $0.093_{-0.002}^{+0.002}$\T\B \\
$\theta\;\mathrm{[deg]}$ & $174.731_{-0.009}^{+0.013}$ & $170.28_{-0.08}^{+0.06}$ & $10.49_{-0.12}^{+0.09}$ & $8.92_{-0.07}^{+0.12}$ & $5.8_{-0.5}^{+3.2}$ & $5.1_{-0.4}^{+171.5}$\T\B \\
$\zeta\;\mathrm{[deg]}$ & $80.63_{-0.02}^{+0.02}$ & $76.37_{-0.09}^{+0.12}$ & $25.3_{-0.1}^{+0.1}$ & $30.4_{-0.2}^{+0.6}$ & $89.3_{-1.2}^{+0.5}$ & $89.2_{-1.3}^{+0.6}$\T\B \\
$\tau\;[-]$ & $1.888_{-0.001}^{+0.002}$ & $2.042_{-0.003}^{+0.002}$ & $2.19_{-0.01}^{+0.01}$ & $2.373_{-0.004}^{+0.003}$ & $1.17_{-0.03}^{+0.06}$ & $1.38_{-0.15}^{+0.05}$\T\B \\
$T_\mathrm{seed}\;\mathrm{[keV]}$ & $0.5113_{-0.0001}^{+0.0003}$ & $0.5522_{-0.0005}^{+0.0006}$ & $0.734_{-0.002}^{+0.002}$ & $0.699_{-0.003}^{+0.014}$ & $0.65_{-0.02}^{+0.06}$ & $0.62_{-0.02}^{+0.02}$\T\B \\
$T_\mathrm{e}\;\mathrm{[keV]}$ & $20.458_{-0.010}^{+0.006}$ & $20.64_{-0.03}^{+0.03}$ & $22.5_{-0.3}^{+0.3}$ & $20.48_{-0.03}^{+0.06}$ & $69.8_{-4.9}^{+5.7}$ & $46.4_{-3.9}^{+5.9}$\T\B \\
$R_\mathrm{in}\;\mathrm{[km]}$ & $39.66_{-0.03}^{+0.02}$ & $39.66_{-0.02}^{+0.02}$ & $39.67_{-0.05}^{+0.03}$ & $39.70_{-0.03}^{+0.01}$ & - & -\T\B \\
$T_\mathrm{in}\;\mathrm{[keV]}$ & $0.10752_{-2e-05}^{+4e-05}$ & $0.11891_{-6e-05}^{+4e-05}$ & $0.11002_{-9e-05}^{+0.00012}$ & $0.1235_{-0.0001}^{+0.0001}$ & - & -\T\B \\
$\mu\;\mathrm{[keV]}$ & - & - & $0.870_{-0.001}^{+0.001}$ & $0.916_{-0.001}^{+0.001}$ & - & -\T\B \\
$\sigma\;\mathrm{[keV]}$ & - & - & $0.196_{-0.001}^{+0.001}$ & $0.181_{-0.002}^{+0.002}$ & - & -\T\B \\
$N_\mathrm{norm}\;\mathrm{[photons/cm^2/s]}$ & - & - & $1.13_{-0.01}^{+0.01}$ & $1.31_{-0.02}^{+0.02}$ & - & -\T\B \\
\hline

    \end{tabular}
\end{table*}

\begin{figure}
    \centering
    \includegraphics[width=\linewidth]{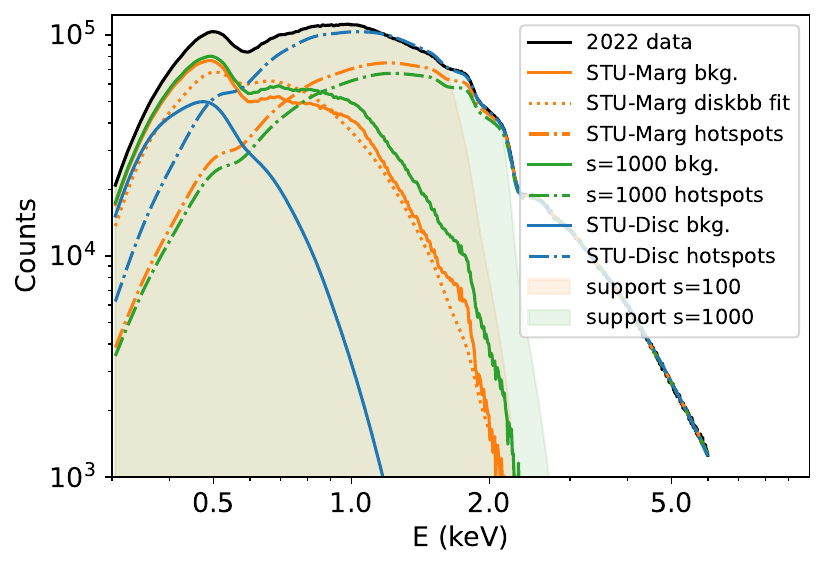}
    \caption{Phase-summed spectra of the data and inferred \stu\ components for the 2022 data. See the caption \Cref{fig:spectra2019} for a description of each component.}
    \label{fig:spectra2022}
\end{figure}

\bsp	
\label{lastpage}
\end{document}